\documentclass[twocolumn,twocolappendix]{aastex63}

\hypersetup{linkcolor=blue,citecolor=blue,filecolor=magenta,urlcolor=cyan}

\shorttitle{Circumgalactic \mgII\ Gas in \EAGLE}
\shortauthors{Ho et al.}

\usepackage{natbib}
\usepackage{bm}
\usepackage{chngcntr}

\begin{document}

%
%
\def\etal{{\rm et al.}}
\def\etali{{\it et al.\thinspace}}
\def\etns{{\rm et\thinspace al.}}   
\def\etaln{et al.\thinspace}

\def\EAGLE{\texttt{EAGLE}}
\def\AREPO{\texttt{AREPO}}
\def\ENZO{\texttt{ENZO}}
\def\GADGET3{\texttt{GADGET-3}}
\def\OWLS{\texttt{OWLS}}
\def\illustris{\texttt{Illustris}}
\def\illustrisTNG{\texttt{IllustrisTNG}}

\def\mgIIdb{\ion{Mg}{2} $\lambda\lambda$2796, 2803}
\def\mgIIdbl{\ion{Mg}{2} $\lambda$2796}
\def\mgIIdbu{\ion{Mg}{2} $\lambda$2803}
\def\oI{[\ion{O}{1}] $\lambda$6300}
\def\sII{[\ion{S}{2}] $\lambda\lambda$6716, 6731}
\def\oIII{[\ion{O}{3}] $\lambda$5007}
\def\hI{\mbox {\ion{H}{1}}}
\def\mgII{\mbox {\ion{Mg}{2}}}
\def\nII{\mbox [{\ion{N}{2}}]}
\def\oVI{\mbox {\ion{O}{6}}}
\def\oVII{\mbox {\ion{O}{7}}}
\def\oVIII{\mbox {\ion{O}{8}}}
\def\cIV{\mbox {\ion{C}{4}}}
\def\halpha{$\mathrm{H}\alpha$}
\def\heII{\mbox {\ion{He}{2}}}

\def\mgplus{$\textrm{Mg}^{+}$}
\def\oVIion{O$^\mathrm{+5}$}

\def\kms{\mbox{km s$^{-1}$}}
\def\kmsMpc{\mbox{km s$^{-1}$ Mpc$^{-1}$}}
\def\kmstb{km s$^{-1}$}
\def\micron{\mbox{$\mu$m}}
\def\modotyr{\mbox {$\rm M_\odot$~yr$^{-1}$}}
\def\msununit{$\rm M_\odot$}
\def\percmsq{\mbox{cm$^{-2}$}}
\def\percmcube{\mbox{cm$^{-3}$}}

\def\rvir{$r_\mathrm{vir}$}
\def\mvir{$M_\mathrm{vir}$}
\def\mstar{$M_\star$}

\def\logmstar{$\log M_\star$}
\def\logmstarmsun{$\log (M_\star/\mathrm{M_\odot})$}

\def\NmgII{$N_\mathrm{MgII}$}

\def\deltavlos{$|\Delta v_\mathrm{LOS}|$}
\def\fmgiimis{$f_\mathrm{MgII,mis}$}
\def\fmgiimisb{$f_\mathrm{MgII,mis}(b)$}
\def\halovc{$v_\mathrm{c,halo}$}
\def\jstar{$\bm{j_\star}$}
\def\jstarscalar{$j_\star$}
\def\fcorotall{$F^\mathrm{corot+det}_\mathrm{all}$}
\def\fcorotdet{$f^\mathrm{corot+det}_\mathrm{det}$}
%
%
%
%
\def \dlow {\mbox{$400 {\rm ~l~mm}^{-1}$}}
\def \dhigh {\mbox{$600 {\rm ~l~mm}^{-1}$}}
\newcommand{\be}{\begin{equation}} \newcommand{\ba}{\begin{eqnarray}}
\newcommand{\ee}{\end{equation}} \newcommand{\ea}{\end{eqnarray}}
\def\-{{\em{---}}}
\def \mA {\mbox{${\rm m \AA} $} }
\def \rr {\mbox{${\rm RR}$} }
\def \rarb {\mbox{${\rm R_AR_B}$} }
\def \rara {\mbox{${\rm R_AR_A}$} }
\def \dd {\mbox{${\rm DD}$} }
\def \dada {\mbox{${\rm D_AD_A}$} }
\def \dadb {\mbox{${\rm D_AD_B}$} }
\def \dr {\mbox{${\rm DR}$} }
\def \darb {\mbox{${\rm D_AR_B}$} }
\def \dara {\mbox{${\rm D_AR_A}$} }
\def \dbra {\mbox{${\rm D_BR_A}$} }
\def \hMpc      {h^{-1}{\rm\ Mpc}}
\def \hkpc      {h^{-1}{\rm\ kpc}}
\def \h         {\hbox{$\, h$} }
\def \hinv      {\hbox{$\, h^{-1}$} }
\def \hinvseven    {\hbox{$\, h_{70}^{-1}$} }
\def\ewr{\mbox {EW$_r$}}
\def\ewo{\mbox {EW$_o$}}
\def\H7{\mbox {$h_{0.7}$}}
\def\naI{\mbox {\ion{Na}{1}}}
\def\mgI{\mbox {\ion{Mg}{1}}}
\def\feI{\mbox {\ion{Fe}{1}}}
\def\znII{\mbox {\ion{Zn}{2}}}
\def\crII{\mbox {\ion{Cr}{2}}}
\def\alI{\mbox {\sc Al~I~}}
\def\alII{\mbox {\sc Al~II~}}
\def\alIII{\mbox {\sc Al~III~}}
\def\mnII{\mbox {\ion{Mn}{2}}}
\def\niII{\mbox {\ion{Ni}{2}}}
\def\feII{\mbox {\ion{Fe}{2}}}
\def\feIII{\mbox {\ion{Fe}{3}}}
\def\sV{\mbox {\ion{S}{5}}}
\def\siIV{\mbox {\ion{Si}{4}}}
\def\siIII{\mbox {\ion{Si}{3}}}
\def\siII{\mbox {\ion{Si}{2}}}
\def\siI{\mbox {\ion{Si}{1}}}
\def\cII{\mbox {\ion{C}{2}}}
\def\cIII{\mbox {\ion{C}{3}}}
\def\llambda{\mbox {$\lambda$}}
\def\hlen{\mbox {$h_{0.7}^{-1}$}}
\def\lstarlya{\mbox {$L^*_{Ly\alpha}$}}
\def\IZw18{I~Zw~18}
\def\m82{M82}
\def\Ab{Abell~}
\def\gi{\mbox {\rm g-i}}
\def\ug{\mbox {\rm u-g}}
\def\br{\mbox {\rm b-r}}
\def\eqn{equation}
\def\vesc{\mbox {$v_{\rm esc}$}}
\def\heha{\mbox {He~I~$\lambda 5876$ / H$\alpha$}}
\def\xhe{\mbox {$\chi({\rm He}) / \chi({\rm H})$} }
\def\he{\mbox {\rm He}}
\def\hii{\mbox {${\rm H}^+$}}
\def\h{\mbox {\rm H}}
\def\mab{\mbox {$\rm m_{AB}$}}
\def\ssp{\baselineskip=13pt plus 1pt minus 1pt}
\def\tsp{\baselineskip=5pt plus 1pt minus 1pt}
%
%
\def\deg{\mbox {$^{\circ}$}}
\def\msun{\mbox {${\rm ~M_\odot}$}}
\def\zsun{\mbox {${\rm ~Z_{\odot}}$}}
\def\lsun{\mbox {${~\rm L_\odot}$}}
\def\msunyr{\mbox {$~{\rm M_\odot}$~yr$^{-1}$}}
\def\angs{\mbox {~\AA}}
\def\lya{\mbox {Ly$\alpha$}}
\def\lyb{\mbox {Ly$\beta$}}
\def\Ha{\mbox {H$\alpha$}}
\def\Hb{\mbox {H$\beta$}}
\def\Hg{\mbox {H$\gamma$}}
\def\tion{\mbox {$T_{\rm ion}$~}}
\def\ch{\mbox {$\bigtriangleup$}}
\def\grad{\mbox {$\bigtriangledown$}}
\def\lstar{\mbox {$L^*$}}
\def\line{\mbox {~$\lambda$}}
\def\lines{\mbox {~$\lambda\lambda$~}}
\def\h0{\mbox {~H$_0$}}
\def\q0{\mbox {~q$_0$}}
%
%
\def\auroral{[OIII]~$\lambda4363$~}
\def\auroral{[OIII]~$\lambda4363$~}
\def\ohsun{\mbox {(O/H)$_{\odot}$~}}

\def\O1ha{[OI]$\lambda6300$~/~H$\alpha$~}
\def\Ru{[OII]$\lambda\lambda3727$~/~[OIII]$\lambda5007$~}
\def\s2ha{[SII]$\lambda\lambda6717,31$~/~H$\alpha$~}
\def\2z2{HeII~$\lambda4686$~}
\def\z7{[NII]~$\lambda6583$ }
\def\N2{[NII]~$\lambda6583$~/~H$\alpha$~}
\def\16z2{[SII]~$\lambda\lambda6717, 6731$ }
\def\HgI{HgI~$\lambda4358$~}
\def\Sdensity{[SII]~$\lambda6717 / \lambda6731$}
\def\Temp{[OIII]~$\lambda\lambda4959 + 5007 ~{\rm to}~ \lambda4363$~}
%
%
\def\j{J}
\def\n{NGC~}
\def\asec{\ifmmode {'' }\else $''~$\fi}  
\def\amin{\ifmmode {' }\else $'~$\fi}    
\def\arcsper{\ifmmode \rlap.{'' }\else $\rlap{.}'' $\fi} 
\def\arcmper{\ifmmode \rlap.{' }\else $\rlap{.}' $\fi} 
\def\sles{\lesssim}
\def\sgreat{\gtrsim}
%
%
\def\gapp{\mbox {$_>\atop{^\sim}$}}  
\def\lapp{\mbox {$_<\atop{^\sim}$}}  
%
\def\kms{\mbox {~km~s$^{-1}$}}
\def\ergsec{~ergs~s$^{-1}$~}
\def\sb{~ergs~s$^{-1}$~cm$^{-2}$~arcsec$^{-2}$}
\def\flux{~ergs~s$^{-1}$~cm$^{-2}$}
\def\flam{~ergs~s$^{-1}$~cm$^{-2}$ \AA$^{-1}$}
\def\cm3{~cm$^{-3}$}
\def\col{\mbox {~cm$^{-2}$}}
\def\mpc3{~Mpc$^{3}$}
\def\mpc-3{~Mpc$^{-3}$}
\def\rate{~sec$~{-1}$}
\def\um{~${\mu}$m~}
\def\fig{{Figure}}
\def\figs{{Figures}}
\def\tbl{{Table}~}
\def\sec{{Sec.}~}
\def\x{{X-ray}~}
\def\xs{{X-rays}~}
\def\X{{X-Ray}~}

%
\def\et{{\rm et\thinspace al.}\ }   
\def\ets{{\rm et\thinspace al.'s}\ }   
\def\reff{\par\noindent\parskip=1pt\hangindent=3pc\hangafter=1}
%
%

%
\def\beginrefs{
         {\normalsize}
         {\noindent}
         \small
        \baselineskip=11pt
        \parindent=0pt
        \frenchspacing
        \parskip=1pt plus 1pt
        \everypar={\hangindent=0.42in}}

\title{Morphological and Rotation Structures of Circumgalactic \mgII\ Gas in the \EAGLE\ Simulation
and the Dependence on Galaxy Properties}


\correspondingauthor{Stephanie Ho}
\email{shho@physics.tamu.edu}

\author[0000-0002-9607-7365]{Stephanie H. Ho}
\affiliation{George P.~and Cynthia Woods Mitchell Institute for Fundamental Physics and Astronomy, Texas A\&M University, College Station, TX 77843-4242, USA}
\affiliation{Department of Physics and Astronomy, Texas A\&M University, College Station, TX 77843-4242, USA}

\author[0000-0001-9189-7818]{Crystal L. Martin}
\affiliation{Department of Physics, University of California, Santa Barbara, CA 93106, USA}

\author[0000-0002-0668-5560]{Joop Schaye}
\affiliation{Leiden Observatory, Leiden University, P.O. Box 9513, 2300 RA, Leiden, The Netherlands}



\begin{abstract}

Low-ionization-state \mgII\ gas 
has been extensively studied in
quasar sightline observations to understand 
the cool, $\sim$$10^4$ K gas in 
the circumgalactic medium.
Motivated by recent observations
showing that the \mgII\ gas around 
low-redshift galaxies
has significant angular momentum,
we use the high-resolution \EAGLE\ 
cosmological simulation
to analyze the morphological and rotation structures
of the $z\approx0.3$ circumgalactic \mgII\ gas
and examine
how they change with the host galaxy properties.
Around star-forming galaxies,
we find that the \mgII\ gas has an 
axisymmetric instead of a spherical distribution,
and the axis of symmetry
aligns with that of the \mgII\ gas rotation.  
A similar rotating structure is less commonly 
found in the small sample of simulated 
quiescent galaxies.
We also examine how often 
\mgII\ gas around galaxies selected
using a line-of-sight velocity cut 
includes gas physically 
outside of the virial radius (\rvir).
For example, 
we show that at an impact parameter of 100 pkpc,
a $\pm500$\kms\ velocity cut
around galaxies with stellar masses of 
$10^9$-$10^{9.5}$ \msununit\ 
($10^{10}$-$10^{10.5}$ \msununit)
selects \mgII\ gas beyond the virial radius
80\% (6\%) of the time.
Because observers typically select \mgII\ gas 
around target galaxies using such a velocity cut,
we discuss how this issue affects the study of
circumgalactic \mgII\ gas properties,
including the detection of corotation.
While the corotating \mgII\ gas 
generally extends beyond 0.5\rvir,
the \mgII\ gas outside of the virial radius 
contaminates the corotation signal and
makes observers less likely
to conclude that gas
at large impact parameters (e.g., $\gtrsim0.25$\rvir)
is corotating.

\end{abstract}


\section{Introduction}
\label{sec:intro}

The reservoir of baryons and metals surrounding galaxies
regulates the interplay between gas accretion and 
feedback of galaxies
and shapes the growth of galactic disks.
Direct imaging of this circumgalactic medium (CGM) 
has proven challenging due to its low gas density.
Observing the circumgalactic gas in absorption 
in the spectra of bright background sources circumvents
this problem and has become a popular
CGM observation approach.
These sightline observations measure 
the absorption lines  
from various ions at different ionization states
and then characterize the CGM properties,
such as the kinematics, radial distribution, 
chemical abundance, and phase structures
(e.g., see \citealt{Tumlinson2017} for a review).

Circumgalactic absorption measurements have
drawn attention to 
the inhomogeneous baryon distribution in the CGM.
Sightlines near the galaxy major or minor axes 
often detect absorption systems 
with large equivalent widths and broad velocity ranges, 
whereas sightlines not aligning with either axis 
rarely detect these strong absorbers
\citep{Bordoloi2011,Bouche2012,Kacprzak2012,
Nielsen2015,Schroetter2019}.  
Such bimodality in spatial geometry 
is frequently observed for low-ionization-state (LIS) 
absorbers (e.g., \mgII),
but it remains controversial
whether the highly ionized \oVI\ absorbers
share the same characteristic \citep{Kacprzak2015}.

In addition to having a non-uniform distribution,
the low-ionization circumgalactic gas 
does not move randomly 
and has significant angular momentum.  
Quasar sightline observations detected Doppler shifts
of the LIS absorption sharing the same sign as the
rotation of the galactic disk,
indicating that the low-ionization CGM
corotates with the galactic disks
\citep{Steidel2002,Kacprzak2010,Kacprzak2011ApJ,
Bouche2013,Bouche2016,Ho2017,Martin2019,Zabl2019}.
The corotation may be unique to the LIS absorbers, 
as the highly ionized \oVIion\ ion
appears to be kinematically uniform and 
does not corotate with the disk 
\citep{Nielsen2017,Kacprzak2019}.

However, revealing the LIS absorbers corotating 
with the galactic disk does not uniquely identify 
the physical structure of the CGM.
\citet{Martin2019} suggested that
the corotating CGM around galaxies with stellar masses 
$\approx 10^{10}$ \msununit\
is likely axisymmetric out to 70-kpc in radius.  
Their measurements showed
a significant drop in \mgII\ covering factor 
for sightlines intersecting
the disk plane at radii larger than 70 kpc,
beyond which the correlation
between the \mgII\ Doppler shift
and the projected rotation velocity on the disk plane
also weakened.
While axisymmetry and rotation together
suggest a rotating disk structure,
a thin disk fails to 
explain the broad linewidth of the \mgII\ absorption
\citep{Steidel2002,Kacprzak2010,Kacprzak2011ApJ,Ho2017}. 
Instead, reproducing the linewidth
requires a thick disk \citep{Steidel2002} or
a combination of the rotation on the extended disk plane
and other components, such as 
outflow and tidal streams \citep{DiamondStanic2016}.
Gas spiraling towards the inner disk presents another
plausible scenario \citep{Ho2017,HoMartin2020}.
Alternatively, numerical simulations show
other features that potentially explain 
the corotating circumgalactic gas observed,
e.g., extended warped gas disks
\citep{Stewart2011ApJ,Stewart2013},
less ``disky'' rotating structures 
\citep{ElBadry2018,Ho2019},
accreting satellites \citep{Shao2018},
infalling streams 
from the cosmic web \citep{Dekel2009,Danovich2015},
and satellite winds \citep{Hafen2018}.  
In fact, decades of absorption-line studies 
have used the kinematic properties
to identify the physical components 
associated with the absorbers
\citep{LanzettaBowen1992,ProchaskaWolfe1997,
CharltonChurchill1998}.
But even now, we have yet to confirm 
the physical structures that correspond to 
individual components of the circumgalactic absorption.

The rotating LIS circumgalactic gas and 
the inhomogeneous distribution of circumgalactic baryons
raise a seemingly simple question: 
what is the general structure of this low-ionization CGM,
i.e., what does the CGM ``look'' like?
For individual objects,
direct imaging of the CGM has only been possible 
for the difficult to interpret Lyman-alpha line, 
see \citet{Cantalupo2014}, \cite{Borisova2016},
and \citet{Cai2019}
for radio quiet quasars and 
\citet{Wisotzki2016,Wisotzki2018} and \citet{Leclercq2017} 
for Lyman-alpha emitters.  
Revealing the faint ionized CGM emission 
in other lines typically
requires stacking many objects 
\citep{Zhang2016,Zhang2018,Guo2020},
but the newly commissioned 
Keck Cosmic Web Imager \citep{Morrissey2018}
has made the imaging of the ionized CGM emission
possible around individual systems;
see the recent \mgII\ emission mappings by
\citet{Burchett2020} of a starburst galaxy merger
and \citet{Chisholm2020} of a
Lyman Continuum emitter.
On the other hand, although sightlines 
around individual typical galaxies probe
the CGM and reveal its properties,
the major limitation of this technique
is the small number of sightlines per galaxy.
Most CGM surveys stack single-sightline observations
to characterize the properties of the average CGM
\citep{Rakic2012,Tumlinson2013,Werk2013,
Turner2014,Borthakur2015,Borthakur2016,Heckman2017,
Chen2018,Rubin2018_i,Martin2019}.  
Only under rare circumstances 
do multi-sightline observations become possible, 
such as with gravitationally lensed quasars 
\citep{Chen2014,Zahedy2016,Rubin2018,Kulkarni2019}
and galaxies \citep{Lopez2018,Lopez2019},
or with multiple bright sources fortuitously 
located behind the target galaxies 
at small projected angular separations
\citep[e.g.,][]{Muzahid2014,Bowen2016,
Peroux2018,Zabl2020}.  
Hence,
CGM tomography remains challenging
until the advent of next generation telescopes.

The uncertainty in associating the absorption system
with the host galaxy presents another challenge
for observational analysis of the CGM.  
Typically, observers associate the absorption system 
with a galaxy 
at small projected separation with the sightline 
and at comparable redshift, 
i.e., with small line-of-sight (LOS) velocity separation.
However, because absorption-line measurements
do not reveal where the absorbing gas 
lies along the sightline, 
the gas potentially resides beyond the CGM 
of the assumed host.  
In addition, 
many faint galaxies may remain undetected.  
These uncertainties lead to possible errors
in determining the circumgalactic gas properties,
especially when individual systems 
cannot be closely examined in surveys 
with thousands of galaxy-absorber pairs
\citep[e.g.,][]{Bordoloi2011,Lan2014,Zhu2014,LanMo2018}.

In contrast to observational studies,
hydrodynamical simulations can directly ``image'' 
the low-density CGM and reveal
the circumgalactic gas distribution and kinematics.  
These simulations generally reproduced
the radial distribution of the column density
of LIS ions 
\citep{Ford2014,Ford2016,Liang2016,
Oppenheimer2018lowion,Nelson2020} 
but underpredicted that of the highly ionized 
\oVIion\ ion
\citep{Hummels2013,Oppenheimer2016,
Gutcke2017,Suresh2017}, 
an issue potentially resolved by black hole feedback
\citep{Nelson2018} 
or fossil AGN proximity zones 
\citep{OppenheimerSchaye2013,Oppenheimer2018agn}.
Simulations also found rotating gas structures
around $z\approx0$ galaxies extended out to 
tens or $\sim100$ kpc \citep{ElBadry2018,Ho2019},
where the angular momentum vector of 
the circumgalactic gas aligned with that 
of the stellar disk \citep{DeFelippis2020,Huscher2020}.
The morphology of the extended gas depends on the 
galaxy properties and the feedback physics
\citep[e.g.,][]{vandeVoortSchaye2012,
Kauffmann2016,Kauffmann2019}.
Nevertheless, these results show that 
the CGM has a rotation component,
agreeing qualitatively with the 
picture of the low-ionization CGM suggested 
by quasar sightline observations.

This paper studies  
the low-ionization circumgalactic gas using the 
high-resolution \EAGLE\ simulation
\citep{Crain2015,Schaye2015}.  
\EAGLE\ has proven capable of broadly
reproducing many galaxy observables, including 
the galaxy stellar mass function \citep{Schaye2015},
the evolution of galaxy masses \citep{Furlong2015}, 
sizes \citep{Furlong2017}, 
colors \citep{Trayford2015,Trayford2017}, 
and gas contents \citep{Lagos2015,Bahe2016,Crain2017}.  
The simulation was not 
calibrated to match observational measurements of
the intergalactic medium (IGM) nor the CGM; 
it was calibrated to match the present-day galaxy 
stellar mass function, the sizes of disk galaxies, 
and the amplitude of the 
galaxy-central black hole mass relation.  
Therefore, \EAGLE\
provides a testbed for understanding 
and testing against the results from CGM observations.
\EAGLE\ shows broad agreements 
with absorption-line statistics 
for \hI\ \citep{Rahmati2015} and 
metal ions
\citep{Rahmati2016,Turner2016,Turner2017,
Oppenheimer2018lowion}.  
In particular, 
\citet{Oppenheimer2018lowion} demonstrated that
\EAGLE\ reproduces the commonly observed
anticorrelation between 
covering fraction of low ions 
(e.g., \siII, \siIII, \cII)
and impact parameter.
The cumulative distribution functions
of the simulated column densites of low ions
match with those from the COS-Halos survey
\citep{Tumlinson2013},
an HST/COS program that characterizes the CGM of 
$z\approx0.2$, $\sim$$L^*$ galaxies
through quasar sightline observations.
The ion ratios and 
pressures of the low-ion metal clumps in \EAGLE\
also agree with that deduced 
from the COS-Halos sample.
Although \EAGLE\ underproduces the \oVIion\ ion,
it reproduces the observed \oVI\ bimodality 
around blue and red galaxies \citep{Oppenheimer2016}.
Not only does \EAGLE\ provide insights on 
interpretating CGM observations and 
on understanding the origin and distribution
of the multiphase gas
\citep[e.g.,][etc.]{Stevens2017,Correa2018a,Correa2018b,
Oppenheimer2018agn,Oppenheimer2018hse,Ho2019,Huscher2020},
but \EAGLE\ also makes predictions for CGM/IGM observations
with future instruments, 
e.g., the column density and equivalent width distribution
of \oVII, \oVIII, and \ion{Ne}{9} absorption systems 
in X-ray observations \citep{Wijers2019,Wijers2020}.

This study examines the LIS \mgII\ gas in 
the low-redshift CGM
and focuses on the morphology and 
the rotation structure.  
We also investigate how often 
selecting the \mgII\ gas around galaxies
using a LOS velocity cut
actually detects \mgII\ outside of
the halo virial radius \rvir;
we refer to this \mgII\ outside of \rvir\
as being ``mis-assigned'',
which creates the issue of
``\mgII\ host galaxy mis-assignment''.
We examine how this issue
affects circumgalactic gas detection and measurement
in sightline observations.
We note that we use the term
``mis-assigned'' simply 
to mean outside of \rvir\ and does not
necessarily mean outside of the CGM,
because the CGM possibly extends 
beyond \rvir\ \citep{Shull2014}.
We focus on \mgII\ in this paper,
because \mgII\ is the most commonly 
studied ion in low-redshift CGM observations
and also because of the recent results on
the \mgII\ rotation kinematics \citep{Ho2017,Martin2019}.
We present this paper as follows.
Section~\ref{sec:sample_simulation} 
describes the \EAGLE\ galaxy selection.  
Section~\ref{sec:mg2misassign} examines 
how the \mgII\ gas outside of \rvir\
affects the detection of \mgII\ gas
with impact parameters smaller than \rvir\
and addresses the significance 
of host galaxy mis-assignments in observations.
In Section~\ref{sec:morph_kin},
we analyze the morphology and the 
rotation structure
of the \mgII\ gas around galaxies
and examine how they vary
across different galaxy populations.  
We also explore 
how the mis-assigned \mgII\ gas
affects the \mgII\ rotation analyses.
In Section~\ref{sec:discussion}, 
we discuss the implication of our results
and relate them to recent observation and 
simulation analyses.
Finally, we conclude in Section~\ref{sec:conclusion}.
Throughout this paper,
we use the flat $\Lambda$CDM cosmology
with $(\Omega_m,\Omega_\Lambda,h) = (0.307,0.693,0.6777)$
adopted by \EAGLE\ from \citet{Planck2014}.

\section{The \EAGLE\ Simulation and Galaxy Selection}
\label{sec:sample_simulation}

\subsection{Simulation Overview}
\label{ssec:simulation}

The \EAGLE\ simulation suite consists of 
a large number of 
cosmological hydrodynamic simulations 
with different cosmological volumes, resolutions, 
and subgrid physics 
\citep{Schaye2015,Crain2015,McAlpine2016}.
\EAGLE\ was run using a modified version 
of the $N$-Body Tree-PM 
smoothed particle hydrodynamics (SPH) code \GADGET3\
(last described in \citealt{Springel2005})
with a new hydrodynamics solver
\citep{Schaller2015}.
State-of-the-art subgrid models were implemented 
to capture unresolved physics,
including radiative cooling and photoheating, 
star formation, stellar evolution and enrichment, 
stellar feedback, 
and active galactic nuclei feedback and black hole growth.
\citet{Schaye2015} introduced a reference model;
the parameters of the subgrid models 
for energy feedback from stars and accreting black holes 
were calibrated to reproduce 
the galaxy stellar mass function at $z\approx0$
and the sizes of present-day disk galaxies.

\begin{deluxetable}{lll}
\tablecaption{Characteristics of the Recal-L0025N0752 simulation used in this paper.  
\label{tb:simulation}}
\tabletypesize{\footnotesize}
\tablewidth{1\linewidth}
\tablehead{
\colhead{} & \colhead{Simulation Property} & \colhead{Value}
}
\startdata
(1) & Box size $L$ (cMpc)                                             & 25\\
(2) & Number of particles $N$                                        & $752^3$\\
(3) & Initial baryonic particle mass $m_\mathrm{g}$ (\msununit)      & $2.26 \times 10^5$\\
(4) & Dark matter particle mass $m_\mathrm{dm}$ (\msununit)          & $1.21 \times 10^6$\\
(5) & Gravitational softening length $\epsilon_\mathrm{com}$ (ckpc)  & 1.33\\
(6) & Maximum softening length $\epsilon_\mathrm{prop}$ (pkpc)       & 0.35
\enddata
\tablecomments{
(1) Comoving box size. 
(2) Number of dark matter particles (initially there is an equal number of baryonic particles).
(3) Initial baryonic particle mass.
(4) Dark matter particle mass. 
(5) Comoving Plummer-equivalent gravitational softening length. 
(6) Maximum proper softening length.
}
\end{deluxetable}


\EAGLE\ defines galaxies as 
gravitationally bound substructures identified by 
the \texttt{SUBFIND} algorithm 
\citep{Springel2001, Dolag2009}.  
In brief, 
the friends-of-friends (FoF) algorithm
places dark matter particles into the same group
if the particle separation is below 0.2 times
the average particle separation.
Baryons are associated with the same FoF halo 
(if it exists)
as their closest dark matter particle.
In each FoF halo,
\texttt{SUBFIND} defines self-bound overdensities
of particles as subhalos;
each subhalo represents a galaxy.
The central galaxy is defined as
the subhalo with the particle at 
the lowest gravitational potential,
and the remaining subhalos are classified 
as satellite galaxies.

In this study, we focus on 
the simulation Recal-L0025N0752,\footnote{
    The ``Recal'' model was calibrated to 
    the same $z\sim0$ galaxy properties 
    as the reference model, 
    but small changes were made to the 
    stellar and AGN feedback subgrid parameters
    as a consequence of the higher resolution 
    compared to the default resolution runs.
    }
which has a box size of 25 cMpc 
and 8 (2) times better mass (spatial) resolution
than the \EAGLE\ default intermediate resolution runs,
e.g.,  Ref-L0100N1504.  
We summarize the simulation parameters 
in Table~\ref{tb:simulation}.  
We use the particle data output\footnote{
    Particle data from snapshots can be downloaded from
    \texttt{\url{http://icc.dur.ac.uk/Eagle/database.php}}
    }
and focus on galaxies at 
a single ``snapshot'' of $z=0.271$;
this redshift is comparable to the galaxy redshifts 
in recent quasar absorption-line studies that
measure the CGM kinematics of low-redshift galaxies
(e.g., \citealt{Ho2017,Martin2019}).
The Hubble parameter at this redshift is
$H(\mathrm{z=0.271}) = 78.0$ \kmsMpc,
and the box size of 25 cMpc corresponds to 1533\kms.
Because \EAGLE\ applies periodic boundary conditions,
the maximum LOS separation is half of the
25 cMpc box size, i.e., 12.5 cMpc,
which corresponds to a velocity difference
of 767 \kms\ (physical) at $z=0.271$.

\subsection{Galaxy Selection from the EAGLE Simulation}
\label{ssec:sample}

\begin{figure}[htb]
    \centering
    \includegraphics[width=1.0\linewidth]{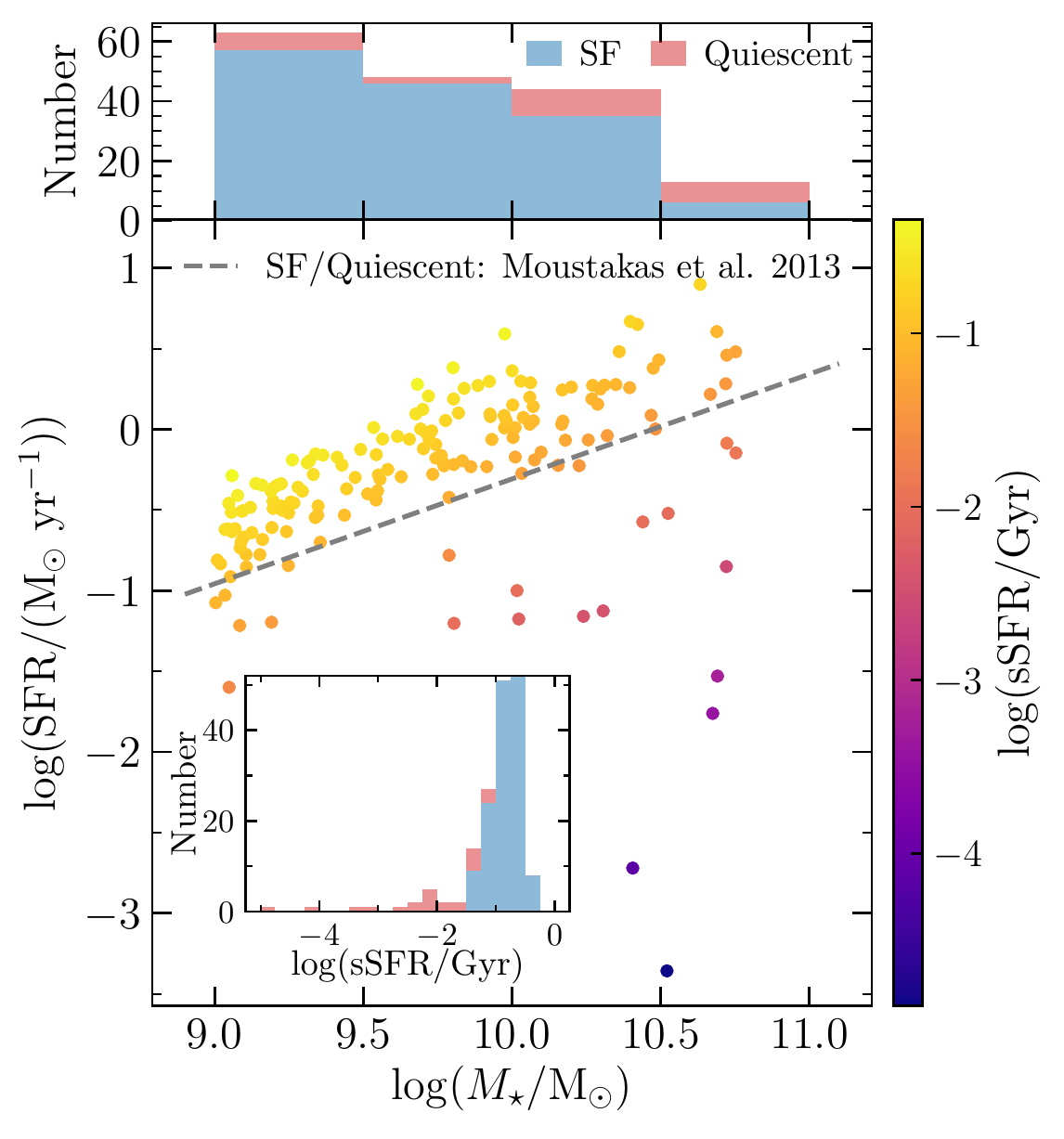}
    \caption{
        Central galaxies on the SFR--\mstar\ plane.  
        Each point is colored by the galaxy 
        sSFR ($ = \mathrm{SFR} / M_\star$).  
        The gray dashed line divides the galaxies 
        into either star-forming or quiescent
        if the galaxies lie above or below the line,
        respectively \citep{Moustakas2013}.  
        Among the 168 central galaxies 
        with stellar masses between 
        $10^9$ and $10^{11}$ \msununit, 
        the sample consists of 144 star-forming
        and 24 quiescent galaxies.
        The histograms at the top and in the inset
        show the distributions of 
        \mstar\ and sSFR, respectively. 
        }
    \label{fig:sfr_mstar} 
\end{figure}

We select central galaxies 
with stellar masses (\mstar)  between 
$10^9$ to $10^{11}$ \msununit.
The stellar mass is 
defined as the total mass of the star particles
associated with the subhalo and 
located within a 30-pkpc radius (in 3D)
from the galaxy center \citep{Schaye2015}.  
The galaxy star formation rate (SFR) is defined
using the same 3D aperture.
Figure~\ref{fig:sfr_mstar} shows the selected galaxies
on the SFR--$M_\star$ plane,
and the color of each point represents 
the galaxy specific SFR (sSFR).
The gray dashed line separates 
star-forming galaxies from quiescent galaxies;
the line is a redshift-dependent relation
fitted from $\sim$120,000 galaxies with 
spectroscopic redshifts from 
the PRism MUlti-object Survey \citep{Moustakas2013}.  
Our sample consists of mainly star-forming galaxies;
the 168 central galaxies include
144 star-forming and 24 quiescent galaxies
(Table~\ref{tb:gal_mstar_count}).
The \mstar\ and the sSFR distributions of the galaxies
are shown in the histograms.

\begin{figure}[htb]
    \centering
    \includegraphics[width=1.0\linewidth]{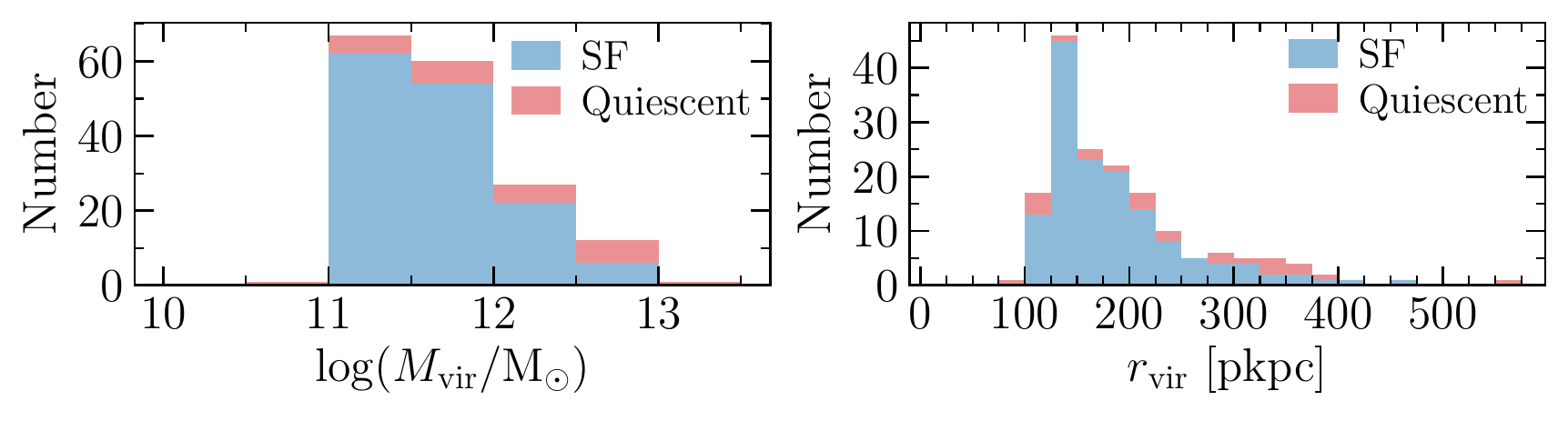}
    \caption{
        Distributions of 
        halo virial mass  \mvir\ and 
        virial radius  \rvir\ 
        of selected central galaxies.
        The virial mass (\textit{left}) 
        ranges from $10^{10.6}$ to $10^{13.2}$ \msununit,
        and the median is $10^{11.6}$ \msununit.
        The median virial radius is 170 pkpc
        and ranges from 78 pkpc to 568 pkpc
        (\textit{right}).
        }
    \label{fig:mhalo_rhalo} 
\end{figure}

\begin{deluxetable}{lcc}
\tablecaption{Galaxy count by stellar mass and 
star-forming vs. quiescent. 
Only central galaxies are included.
\label{tb:gal_mstar_count}}
\tabletypesize{\footnotesize}
\tablewidth{1\linewidth}
\tablehead{
\colhead{} & \colhead{Star-forming} & \colhead{Quiescent}
}
\startdata
$9.0 \leq \log (M_\star/\mathrm{M_\odot}) < 9.5$   & 57 & 6 \\
$9.5 \leq \log (M_\star/\mathrm{M_\odot}) < 10.0$  & 46 & 2 \\
$10.0 \leq \log (M_\star/\mathrm{M_\odot}) < 10.5$ & 35 & 9 \\
$10.5 \leq \log (M_\star/\mathrm{M_\odot}) < 11.0$ & 6  & 7 \\
\hline
Total Number of Galaxies        & 144 & 24
\enddata
\end{deluxetable}


The selected central galaxies span a 
halo virial mass range 
between $10^{10.6}$ and $10^{13.2}$ \msununit\
with a median of $10^{11.6}$ \msununit\
(left panel of Figure~\ref{fig:mhalo_rhalo}).  
We define the virial radius \rvir\ as the 
radius enclosing an average density of
$\Delta_\mathrm{vir} \rho_c(z)$,
where $\rho_c(z)$ represents the critical density 
at redshift $z$, 
and the overdensity $\Delta_\mathrm{vir}$ follows
the top-hat spherical collapse calculation
in \citet{BryanNorman1998}.  
For our galaxy sample at $z=0.271$,
the median virial radius is 170 pkpc,
and the individual virial radii vary from 78 to 568 pkpc
(right panel).

For each galaxy, we define its orientation
using the net specific angular momentum vector of 
the star particles \jstar\  
within the 30-pkpc aperture.  
The plane that intersects the galaxy center
and is normal to the angular momentum vector
defines the disk midplane.
We use this orientation to define 
how we project galaxies onto 2D planes 
while observing the galaxies at fixed inclination angles.

\subsection{Two-dimensional Projection Maps}
\label{ssec:2dmap}

We project galaxies either along 
fixed simulation box axes 
(Section~\ref{sec:mg2misassign})
or at fixed galaxy inclination angles
(Section~\ref{sec:morph_kin}).  
Then, we produce the \mgII\ column density maps and
the \mgII-weighted LOS velocity maps.  
The calculation of the ionic column density
requires the tracked element abundance
and the ion fraction, which is 
the number of atoms in each ionization state
relative to the total number of atoms 
of the element in the gas phase.
We obtain the ion fraction ($f_\mathrm{ion}$)
using the tables from the fiducial model presented in
\citet{PloeckingerSchaye2020}, 
\texttt{UVB\_dust1\_CR1\_G1\_shield1}.\footnote{
    The hdf5 tables are publicly available on
    \url{http://radcool.strw.leidenuniv.nl}
    and 
    \url{https://www.sylviaploeckinger.com/radcool}.}
They used \texttt{CLOUDY v17.01} \citep{Ferland2017}
to tabulate the properties of gas 
(e.g., cooling and heating rates, ion fraction, etc.)
for a wide range of gas density, temperature,
metallicity, and redshift.
In this fiducial model,
the calculations assume ionization equilibrium,
and the gas is exposed to the
redshift-dependent UV/X-ray background
by \citet{FaucherGiguere2020}\footnote{
    \citet{PloeckingerSchaye2020} 
    modified the $z>3$ UV/X-ray background in
    \citet{FaucherGiguere2020}  
    to make the treatment of 
    attenuation before \hI\ and \heII\
    reionization more self-consistent 
    (see their Appendix B).
    This modification is irrelevant to this work
    at $z=0.271$.},
the interstellar radiation field,
and cosmic rays.  
The model also accounts
for the depletion of metals onto dust grains,
tabulated as the number fraction of atoms
depleted ($f_\mathrm{dust}$) for each element.
The effect of self-shielding is included;
the radiation at the center
of a gas cloud is attenuated by its dust and gas
and can be self-shielded from photoionizing radiation,
leaving the cloud cold and neutral
at the inside but ionized at the outside.
This is modeled by passing the incident radiation field
through a gas shielding column,
which is set to half of the local Jeans column density;
the latter is the typical scale for self-gravitating gas.
Using the $f_\mathrm{ion}$ and $f_\mathrm{dust}$ tables,
we obtain the ion balances for each SPH particle
and calculate the number of ions 
through a column in the simulation box
(see Section 2 of \citealt{Wijers2019} for details
of creating column density maps from 
SPH particles).
Note that the ion fraction will differ somewhat 
from that used to compute cooling rates 
during the simulation, since \EAGLE\ used 
an older version of \texttt{CLOUDY}, 
a different UV background model 
and did not include self-shielding.  
We do not expect this to be important, however, 
because magnesium is not an important coolant
\citep{Wiersma2009}.

In addition, due to 
the lack of resolution to resolve 
the interstellar gas phase at $\ll10^4$ K,
\EAGLE\ imposes a temperature floor, 
such that the effective equation of state 
prevents artificial Jeans fragmentation
\citep{Schaye2008}.
Therefore, before calculating the \mgplus\ ion fraction,
we change the temperature of star-forming gas 
to $10^4$ K, which is the typical temperature 
of the warm-neutral interstellar medium (ISM).

We select the \mgII\ gas using two separate ways
to make the projection maps.
The first method is to include only the gas within \rvir.
This method excludes the \mgII\ gas
physically separated from the target galaxies 
but appearing to be closeby on the projection maps.
As for the second method, 
we include the \mgII\ gas within a certain
LOS velocity separation \deltavlos\
from the systemic velocity of the target galaxy.  
We adopt $|\Delta v_\mathrm{LOS}| = 500$\kms,
which is commonly used in observational studies
to associate absorption systems with host galaxies
\citep[e.g.,][]{Chen2010,Chen2018,Werk2013}.

We produce the maps of the \mgII\ column density
and the \mgII-weighted LOS velocity
using the selected \mgII\ gas.
Each pixel on the column density maps shows 
the column summed along the pathlength 
enclosed by \rvir\ or the \deltavlos\ window,
and the LOS velocity maps show the projected velocity 
weighted by the column density of the enclosed gas.  
Each pixel has
an area of either (1.25 pkpc)$^2$ or (0.005 \rvir)$^2$. 
While varying the pixel size changes 
the column density at the pixel,
e.g., a coarser pixel smoothes out the 
high column density region,
the conclusions of our analyses remain unchanged
if we double or halve the pixel size
(see also the discussion in the 
Appendix of \citealt{Wijers2019}).
While we adopt the fiducial model in 
\citet{PloeckingerSchaye2020} 
throughout this paper, 
we have explored using their other models
to understand 
how interstellar radiation and cosmic rays,
self-shielding, and dust depletion
affect the CGM \mgII\ column density.
We find that the effect of dust depletion is tiny
and changes the \mgII\ column density 
by no more than 0.02 dex.
Turning off the interstellar radiation and cosmic rays 
(model \texttt{UVB\_dust1\_CR0\_G0\_shield1})
increases the \mgII\ column density 
typically by 0.01 to 0.02 dex;
this small change suggests that most of 
the circumgalactic \mgII\
does not come from ISM densities.
Lastly, comparing the results from the 
fiducial model to that of without self-shielding 
(model \texttt{UVB\_dust1\_CR1\_G1\_shield0})
shows that self-shielding 
boosts the \mgII\ column density 
typically by 0.1 to 0.3 dex.
This suggests that including self-shielding
is important for analyzing \mgII\ gas, 
especially because
\mgII\ traces higher densities 
compared to other commonly observed low ions, 
such as \siII\ and \cII.\footnote{
    \mgII\ traces gas with 
    $n_\mathrm{H} \gtrsim 10^{-2}$ \percmsq,
    whereas \siII\ and \cII\ trace
    $n_\mathrm{H} \gtrsim 10^{-3}$ \percmsq\
    and $n_\mathrm{H}\gtrsim 10^{-4} $ \percmsq\
    gas, respectively \citep[e.g.,][]{Tumlinson2017}.
    }

To resemble the ``detection limit'' 
of \mgII\ absorption in CGM observations, 
we impose a \mgII\ detection limit of 
\NmgII\ = $10^{11.5}$ \percmsq.  
This limit is comparable to 
the typical observational limit of $\sim 10^{12}$ \percmsq
\citep[e.g.,][]{Werk2013,Martin2019},
and we also take into account
the potentially $\approx 0.3$ dex 
too low magnesium nucleosynthetic yields 
in \EAGLE\ \citep{Segers2016}.
Hence, we consider the \mgII\ gas as ``detectable''
only if its column density exceeds the detection limit 
of \NmgII\ = $10^{11.5}$ \percmsq.

\section{\mgII\ Host Galaxy Mis-assignment}
\label{sec:mg2misassign}

Observers typically associate a \mgII\ absorption system
with a galaxy host 
close to the sightline
\citep[typically $< 300$ kpc, e.g.,][]{Churchill1996,
Kacprzak2007,Nielsen2013_i}
and with redshift
similar to that of the absorption.
The latter is typically defined
using a LOS velocity window \deltavlos\ 
centered at the galaxy systemic redshift
\citep[e.g., 500\kms,][]{Chen2010,Chen2018,Werk2013}.\footnote{
    In \citet{Chen2010} and \citet{Chen2018}, 
    even though
    they used a LOS velocity 
    search window of $\pm500$\kms, 
    over 90\% of the absorbers are found 
    to be within 300\kms\ of the galaxies.
    } 
However, observations cannot determine 
where the absorbing gas lies along the sightline;
the absorbing gas potentially lies beyond the
virial radius of the galaxy
but satisfies the LOS velocity selection criterion,
i.e., the \mgII\ gas is mis-assigned.
In this section, we address this issue of 
\mgII\ host galaxy mis-assignment
and explore how this problem varies
across galaxies with different characteristics.

\subsection{Detecting \mgII\ Gas Mis-assigned 
to the Target Galaxies}
\label{ssec:det_misassign}

\begin{figure*}[bht]
    \centering
    \includegraphics[width=0.95\linewidth]{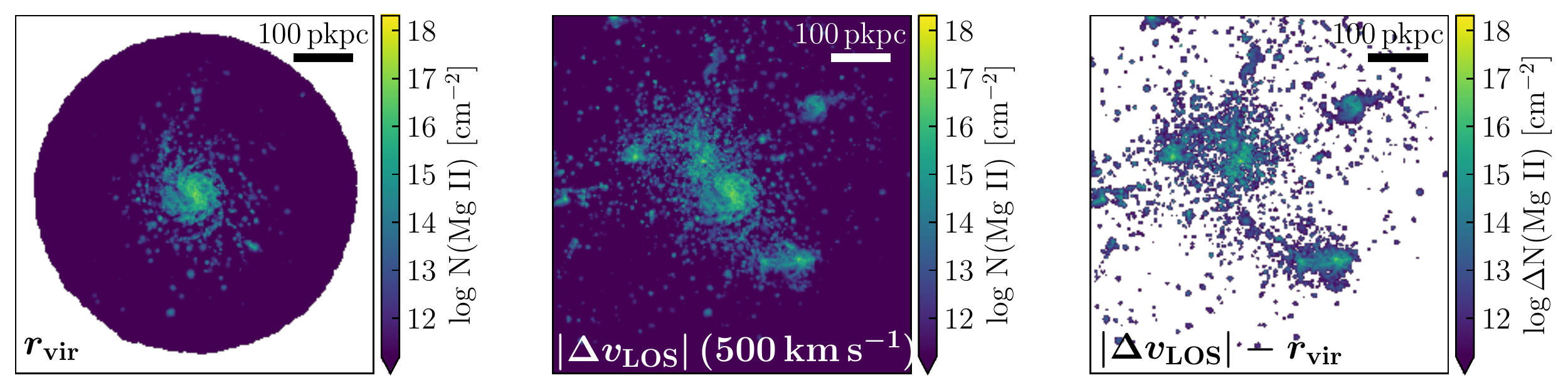}
    \caption{
        Example of mis-assigning \mgII\ gas structures
        to a star-forming galaxy with 
        \logmstarmsun\ $=10.4$.
        \textit{(Left)} The \mgII\ gas within \rvir.
        The region outside of \rvir\ 
        is shown in white.
        \textit{(Middle)} The \mgII\ gas 
        within the LOS velocity window of 
        $\pm 500$\kms\ from the galaxy systemic velocity.
        \textit{(Right)} 
        Column density difference between the \rvir\ 
        and the \deltavlos\ columns,
        i.e., 
        $\log (N_\mathrm{MgII,|\Delta v|} - N_\mathrm{MgII,rvir})$.
        Below the detection limit of $10^{11.5}$ \percmsq,
        the excess \mgII\ column density 
        is not ``detectable''
        and is shown in white.
        The $x$ and $y$ axes of the maps align with 
        those of the simulation box, 
        and the column density is integrated 
        along the $z$-axis.
        All maps are 600 pkpc $\times$ 600 pkpc$^2$.  
        }
    \label{fig:misassign_example}
\end{figure*}

For each galaxy, 
we integrate the column along 
the $z$-axis of the simulation box
and produce two sets of \mgII\ column density maps. 
The first map includes only the \mgII\ gas within \rvir,
whereas the second map 
includes the \mgII\ gas within
\deltavlos\ $= 500$\kms\ 
from the galaxy systemic velocity.
We name these two columns the \rvir\ column 
and the \deltavlos\ column, respectively,
and calculate their column density difference, 
i.e., $\log \Delta N_\mathrm{MgII} = 
\log (N_\mathrm{MgII,|\Delta v|} - N_\mathrm{MgII,rvir})$.
This difference
represents the column density of the \mgII\ gas 
outside of \rvir\ but would have 
erroneously associated with the galaxy if we 
use the \deltavlos\ window to identify the \mgII\ host.
This mis-assigned \mgII\ gas 
is ``detectable'' if its column density 
($\log \Delta N_\mathrm{MgII}$) exceeds 
the \mgII\ detection limit.

The column density maps in Figure~\ref{fig:misassign_example} 
illustrate an example of detecting 
the mis-assigned \mgII\ gas.  
When we include the \mgII\ gas within the 
$\pm 500$\kms\ LOS velocity window (middle),
the map shows additional regions with
high \mgII\ column density
compared to that of the \mgII\ gas within \rvir\ (left).
The column densities of these extra \mgII\ structures 
exceed the \mgII\ detection limit (right),
and sightline observations 
would have associated these structures with 
the target galaxy.

\subsection{The Significance of Detecting
Mis-assigned \mgII\ Gas}
\label{ssec:f_misassign}

We examine how the \mgII\ gas outside of \rvir\
affects the \mgII\ detection rate around target galaxies.
For individual galaxies, 
we produce the maps of the column density difference 
($\log \Delta N_\mathrm{MgII}$)
between the \rvir\ and the \deltavlos\ columns
as in Figure~\ref{fig:misassign_example}.  
We flag the pixels as detecting mis-assigned \mgII\ gas
if $\log \Delta N_\mathrm{MgII}$ 
exceeds the \mgII\ detection limit.\footnote{
    This $\log \Delta N_\mathrm{MgII}$  calculation
    implicity assumes that 
    the \mgII\ gas within the \rvir\ column
    is also within the \deltavlos\ $=500$\kms\ column.  
    This is a reasonable assumption;
    absorption-line studies showed that 
    the circumgalactic gas is bound to the galaxies
    \citep{Werk2013}.
    We analyzed the simulation and 
    also found that this assumption 
    has negligible effects 
    on the \mgII\ mis-assignment fractions
    we are calculating.
    }
Then, we stack these maps
for galaxies in different stellar mass bins.
At each pixel of individual stacks,
we calculate two quantities: 
(1) the number of galaxies flagged,
and
(2) the number of galaxies with the \deltavlos\ column 
exceeding the \mgII\ detection limit,
i.e., the \mgII\ gas is ``detectable'' in the first place.
We bin the pixels by every 10 pkpc in impact parameter $b$,
which is the projected separation between 
individual pixels and the galaxy center. 
Then, we divide the two quantities (after binning) 
to obtain the \mgII\ mis-assignment fraction \fmgiimis.
In other words, 
we are asking the question: 
for a LOS with a detectable amount of \mgII\ at 
\deltavlos\ $\leq 500$\kms, 
what is the probability that a detectable fraction
of this \mgII\ resides beyond \rvir?

\begin{deluxetable}{llr}
\tablecaption{Fitted parameters for the \mgII\ mis-assignment fraction represented by $f_\mathrm{MgII,mis}(b) = 1/ (1 + e^{-\beta (b - b_\mathrm{1/2})})$.
\label{tb:fitparam_fmis}  }
\tabletypesize{\footnotesize}
\tablewidth{1\linewidth}
\tablehead{
\colhead{Galaxy Stellar Mass} & \colhead{$\beta$ (pkpc$^{-1}$)} & \colhead{$b_\mathrm{1/2}$ (pkpc)}
}
\startdata
$9.0 \leq $ \logmstarmsun\ $< 9.5$  & $0.0536 \pm 0.0024$ & $78 \pm 1$ \\
$9.5 \leq $ \logmstarmsun\ $< 10.0$  & $0.0393 \pm 0.0011$ & $148 \pm 1$ \\
$10.0 \leq $ \logmstarmsun\ $< 10.5$  & $0.0254 \pm 0.0004$ & $214 \pm 1$ \\
$10.5 \leq $ \logmstarmsun\ $< 11.0$  & $0.0131 \pm 0.0005$ & $321 \pm 3$ \\
\enddata
\end{deluxetable}
%

\begin{figure}[thb]
    \centering
    \includegraphics[width=0.95\linewidth]{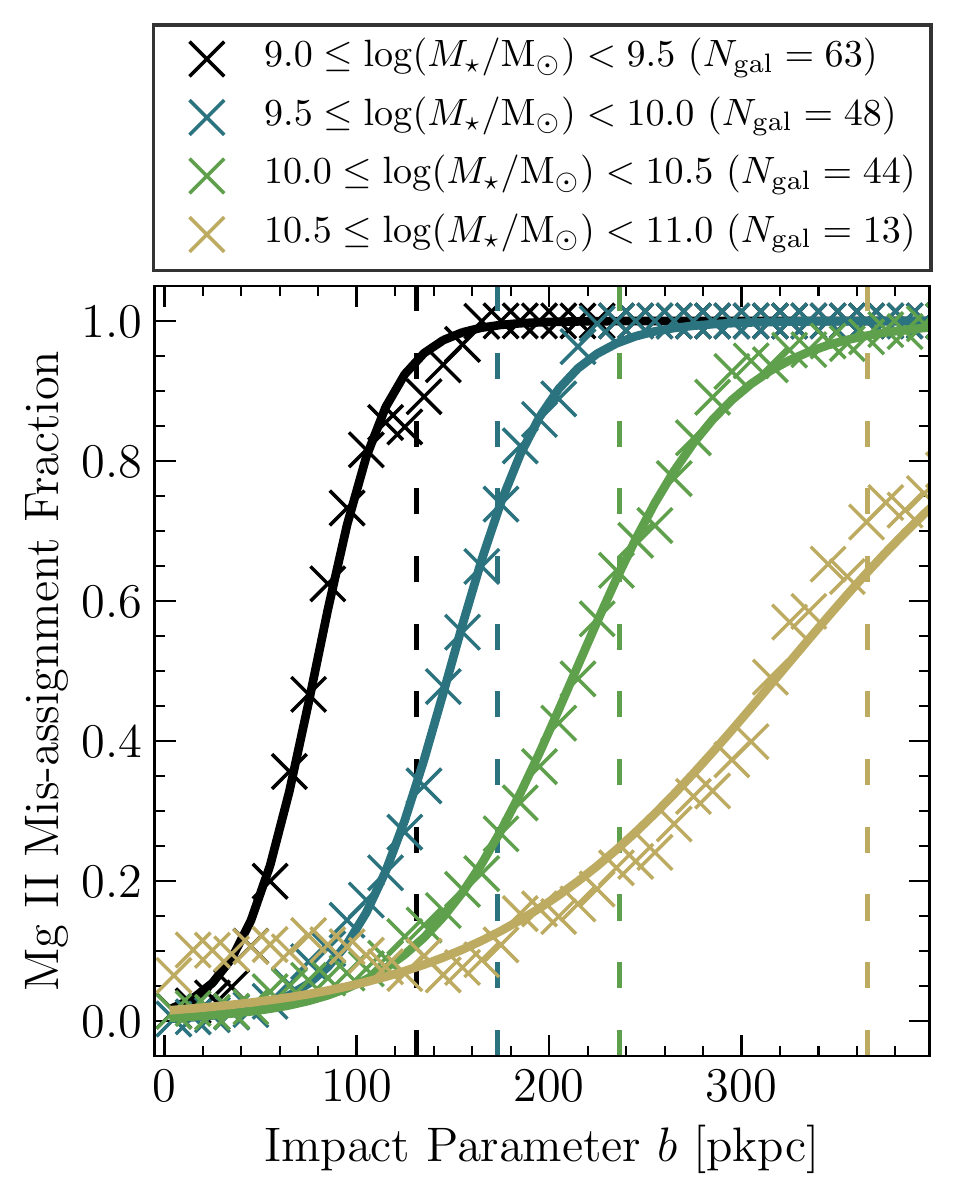}
    \caption{
        Variations of 
        \mgII\ mis-assignment fractions 
        with impact parameter and 
        galaxy stellar mass.
        The mis-assignment fraction \fmgiimis\ measures 
        how likely it is that a detectable amount of
        \mgII\ gas lies outside of \rvir\ but 
        is selected by the \deltavlos\ $=500$\kms\ window,
        i.e., the \mgII\ gas is mis-assigned
        to the target galaxy.
        Each color represents the result
        from each stack of galaxies
        in each stellar mass bin.
        The vertical lines show the median \rvir\
        of the corresponding galaxy stacks.
        Each curve shows the analytic fit 
        to \fmgiimis\ for each galaxy stack.
        }
    \label{fig:mg2_misassignment} 
\end{figure}

Figure~\ref{fig:mg2_misassignment} plots 
\mgII\ mis-assignment fraction, \fmgiimis, 
as a function of impact parameter $b$ 
and shows how this changes with 
galaxy stellar mass.
The four colors, from dark to light,
show the results from the stacks of galaxies
with increasing stellar masses.
Clearly, 
the \mgII\ mis-assignment fraction increases 
with impact parameter and approaches 1.  
The latter is expected;
when  $b > $ \rvir\ for individual galaxies,
the column no longer intersects any region 
within \rvir\ (vertical lines show the medians),
and any \mgII\ gas detected must be 
outside of \rvir.\footnote{
    The $f_\mathrm{MgII,mis}$ does not always 
    reach 1 at the vertical lines of
    Figure~\ref{fig:mg2_misassignment},
    because the vertical lines only show 
    the median \rvir\ of the galaxies in each stack.
    }

We fit an analytical function to describe 
the increase of the \mgII\ mis-assignment fraction 
with impact parameter.
We adopt the functional form of
\fmgiimisb$ = 1/ (1 + e^{-\beta (b - b_\mathrm{1/2})})$,
where $\beta$ describes the steepness of the rise,
and $b_\mathrm{1/2}$ represents the impact parameter where
\fmgiimis$ = 0.5$.  
The solid curves in Figure~\ref{fig:mg2_misassignment}
show the fits, 
and Table~\ref{tb:fitparam_fmis} lists the 
best-fit $\beta$ and $b_\mathrm{1/2}$ for the 
stacks of galaxies with 
different stellar masses.
For example, 
$b_\mathrm{1/2}$ is only 78 pkpc 
for galaxies at the lowest stellar mass bin.
This means that 50\% of the time, 
sightline observations with impact parameters of 
$\approx100$ pkpc would mis-assign the galaxy host 
of the ``detected'' \mgII\ gas.  
The best-fit
$b_\mathrm{1/2}$ ($\beta$) increases (decreases)
with increasing galaxy stellar mass.  
Hence, at a fixed impact parameter,
the detected \mgII\ gas around a less massive galaxy
is more likely to be mis-assigned.

Our calculated 
\mgII\ mis-assignment fraction
represents a conservative estimate.
As a result of the 
\EAGLE\ periodic boundary conditions,
the maximum LOS separation in the 25 cMpc box
at $z=0.271$ is 767 \kms\ 
(see Section~\ref{ssec:simulation}).
Because this is not significantly larger than 
our \deltavlos\ $\leq 500$\kms\ selection,
it is possible that we underestimate the 
level of contamination in the 
$\pm$500\kms\ window.
In particular, the 25 cMpc box is too small 
to contain massive clusters,
for which the galaxy peculiar velocities 
possibly reach $\sim$1000 \kms.
In addition,
the \mgII\ gas could still be 
``mis-assigned'' even if it resides within \rvir\
of the assumed host, because
the \mgII\ could arise in an undetected 
satellite galaxy.
Therefore, the \mgII\ mis-assignment problem 
can be even worse in observational analyses.

From the observers' perspective, 
whether the mis-assigned \mgII\ gas ``detected'' 
in a column (analogous to a sightline) significantly
affects the \mgII\ measurements,
e.g., column density and velocity dispersion, 
depends on the relative column density difference
between the mis-assigned \mgII\ gas 
and the \mgII\ gas within \rvir.
For example, 
if the column density of the \mgII\ gas inside of \rvir\
is orders-of-magnitude higher than that 
of the mis-assigned \mgII\ gas,
then the mis-assigned \mgII\ gas will increase the 
overall velocity spread but have negligible effect 
on the total column density measured.
Therefore, we repeat the calculation of the
\mgII\ mis-assignment fraction 
by adding column density ratio requirements
while flagging the ``detected'' mis-assigned \mgII\ gas;
we require the column density of 
the mis-assigned \mgII\ gas to reach 
either at least 10\% or 100\% of
that of the \mgII\ gas within \rvir.
However, we find that
adding either requirement only decreases
the \mgII\ mis-assignment fractions
by no more than 0.03 in magnitude 
compared to those in Figure~\ref{fig:mg2_misassignment}.
This suggests that a single column
rarely intersects high density \mgII\ gas 
both within and outside of \rvir.
The major observational consequence of the 
mis-assigned \mgII\ gas is the increase in 
the \mgII\ detection rate,
i.e., the binary classification of 
detection vs.~non-detection,
rather than increasing the column density
and/or velocity spread in individual
columns (sightlines)
that already detect \mgII\ gas from inside \rvir.

\section{Morphology and Rotation of the \mgII\ Gas}
\label{sec:morph_kin}

\begin{figure}[thb]
    \centering
    \includegraphics[width=1.0\linewidth]{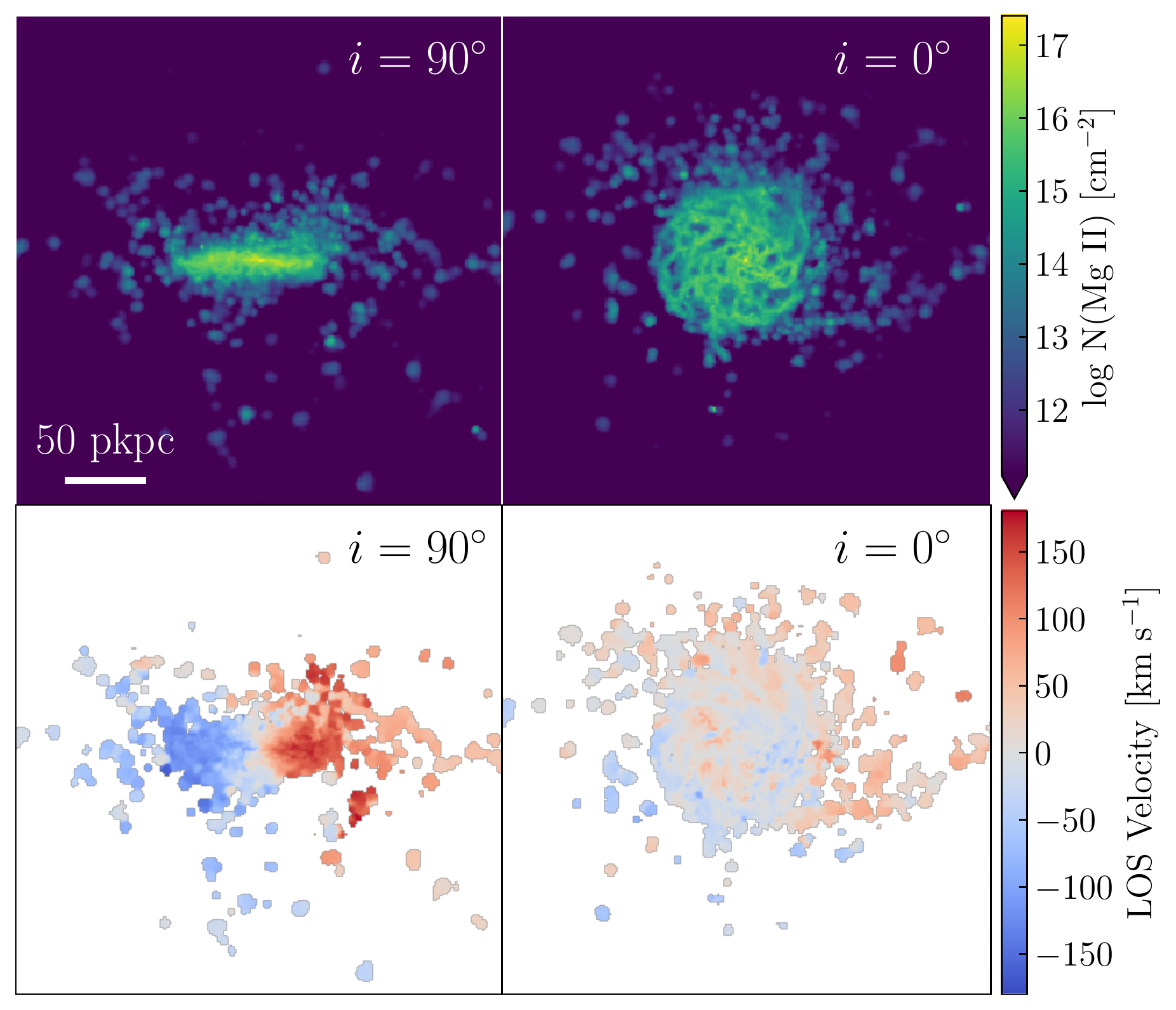}
    \caption{
        \mgII\ column density and LOS velocity maps
        of a star-forming galaxy of 
        \logmstarmsun\ $= 9.93$.
        This galaxy is projected 
        at $i=90$\deg\ \textit{(left)} 
        and 0\deg\ \textit{(right)}.
        The top and bottom rows show 
        the \mgII\ column density and 
        the \mgII-column-density-weighted LOS velocity,
        respectively.  
        A positive (negative) LOS velocity indicates 
        the gas is receding (approaching).
        These projection maps illustrate that 
        the \mgII\ gas is 
        morphologically and kinematically ``disky''.
        Regions with \mgII\ gas below 
        the detection limit of
        $N_\mathrm{MgII} = 10^{11.5}$\percmsq\
        are shown in purple and white on the 
        column density and LOS velocity maps,
        respectively.
        Only the \mgII\ gas within \rvir\ of the
        galaxy is included.
        All maps have the same scale  
        with $300$ pkpc on each side.
        }
    \label{fig:pjdemo} 
\end{figure}

Although quasar sightline observations
reveal the inhomogeneous distribution of the \mgII\ gas
\citep{Bouche2012,Kacprzak2012}
and the corotation between the \mgII\ gas 
and galactic disks
\citep{Ho2017,Martin2019,Zabl2019}, 
these observations do not uniquely identify 
the \mgII\ morphological structure
nor the extent of the corotation. 
For example, is the \mgII\ gas isotropically distributed 
around the galaxy, or does the gas resemble 
a disk structure?
In this section, we examine the distribution 
of the corotating \mgII\ gas
and study its morphological and rotation structures 
around galaxies.  
We analyze how these results vary
with galaxy properties,
and we also discuss
how the mis-assigned \mgII\ gas affects our analysis.


We make 2D projection maps
of the \mgII\ column density
and the LOS velocity 
(weighted by the \mgII\ column density)
to study the \mgII\ morphology and kinematics.
We use the net specific angular momentum vector of 
the star particles \jstar\  to define 
how we project the galaxies (Section~\ref{ssec:sample}).
For each galaxy, we orient \jstar\  such that 
it has components along the $+y$ (upward on the 2D map)
and the $-z$ axes (into the 2D map)
but not along the $x$-axis.
This orientation makes the  $+x$-direction 
the receding side of the net rotation.
We project each galaxy at fixed inclination angles of  
$i = $ 90\deg, 60\deg, 30\deg, and 0\deg,
where $i$ is the angle between \jstar\  and 
the $-z$-axis of the 2D map.  
For example, for the $i = $ 0\deg\ (90\deg) projection,
\jstar\  points along the $-z$-axis ($+y$-axis).

Figure~\ref{fig:pjdemo} illustrates 
the 2D projections of a star-forming galaxy with
\mstar\ $\approx 10^{10}$ \msununit.
The \mgII\ column density maps (top row)
at $i=90$\deg\ and 0\deg\ 
show that the \mgII\ distribution resembles
that of an axis-symmetric disk viewed 
at edge-on and face-on, respectively.  
The bottom left panel shows the $i=90$\deg\ projection  
of the \mgII\ LOS velocity.
Not only do the blueshifted and redshifted sides 
indicate rotation,
but the sense of rotation also follows that 
of the stellar component of the galaxy, 
i.e., the $x>0$ side is redshifted and thereby receding.
In particular, within the radius of around 50 pkpc,
the rotation signature largely disappears 
at $i=0$\deg\ (bottom right),
an unsurprising outcome from
the projection effect of disk rotation.
A weak signature of rotation is still visible 
especially at large radii, however, 
suggesting that how well the \mgII\ gas corotates 
with the stellar component of the galaxy depends on radius.
Nevertheless, from both morphological and 
kinematic perspectives,
this example galaxy has a disk-like \mgII\ structure.

\subsection{Morphological Structure of \mgII\ Gas}
\label{ssec:morphology}

\begin{figure*}[thb]
    \centering
    \includegraphics[width=0.95\linewidth]{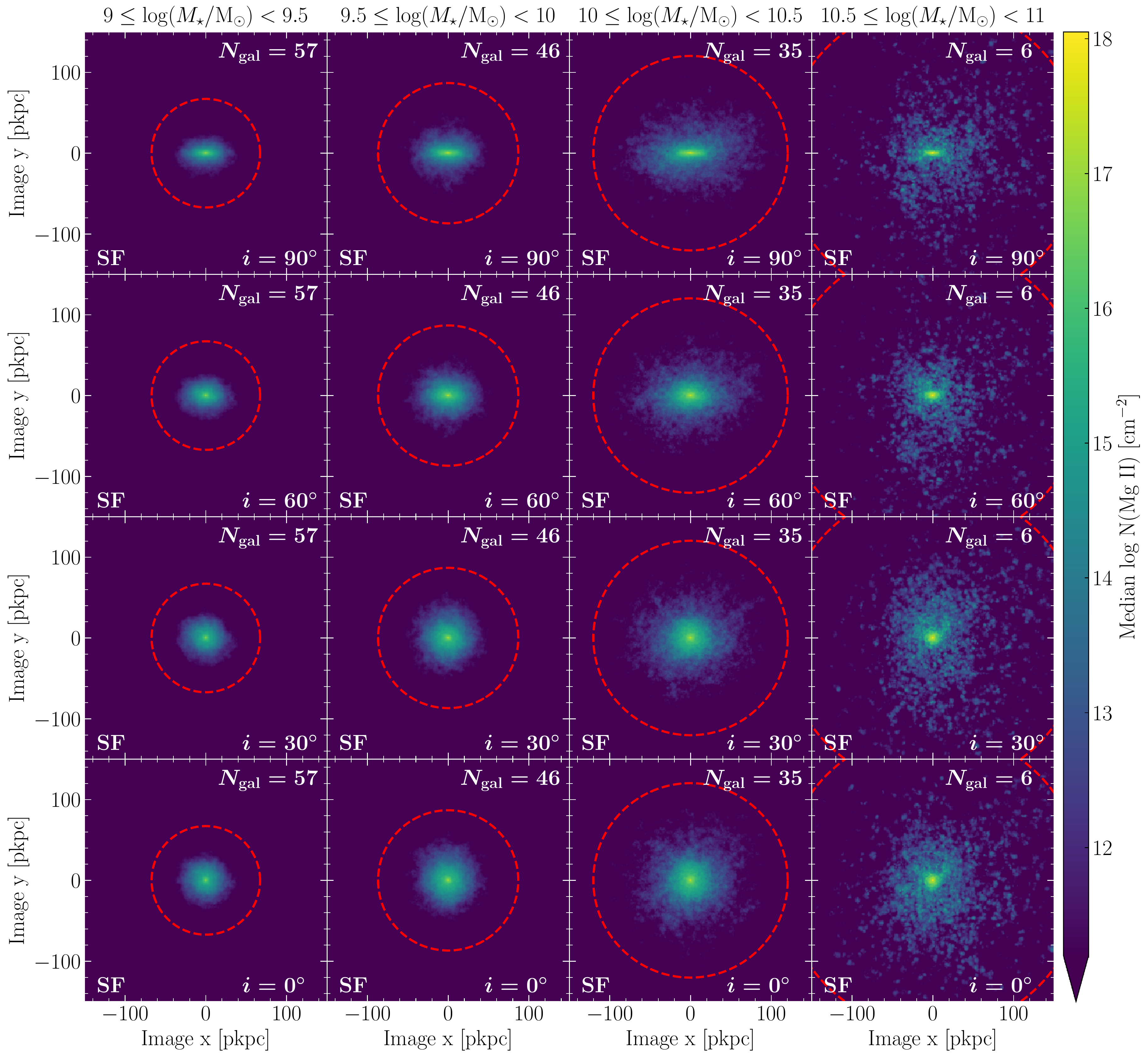}
    \caption{
        Median \mgII\ column density
        around star-forming galaxies
        projected at different inclination angles
        and grouped into different stellar mass bins.
        Only the gas within \rvir\ of 
        individual galaxies is included.
        From left to right, 
        each column shows the result for galaxies
        with increasing stellar masses
        (labeled at the top of each column).
        Each row represents galaxies projected
        at different inclination angles $i$
        before stacking
        (labeled at the lower right of each panel).
        The number of galaxies $N_\mathrm{gal}$
        in each stack
        is labeled at the upper right.
        Each red dashed circle shows the 
        median 0.5\rvir\ of the galaxies in the stack.
        The \mgII\ gas is clearly 
        not spherically distributed.
        }
    \label{fig:mg2logN_sf_16p} 
\end{figure*}

\begin{figure*}[thb]
    \centering
    \includegraphics[width=0.95\linewidth]{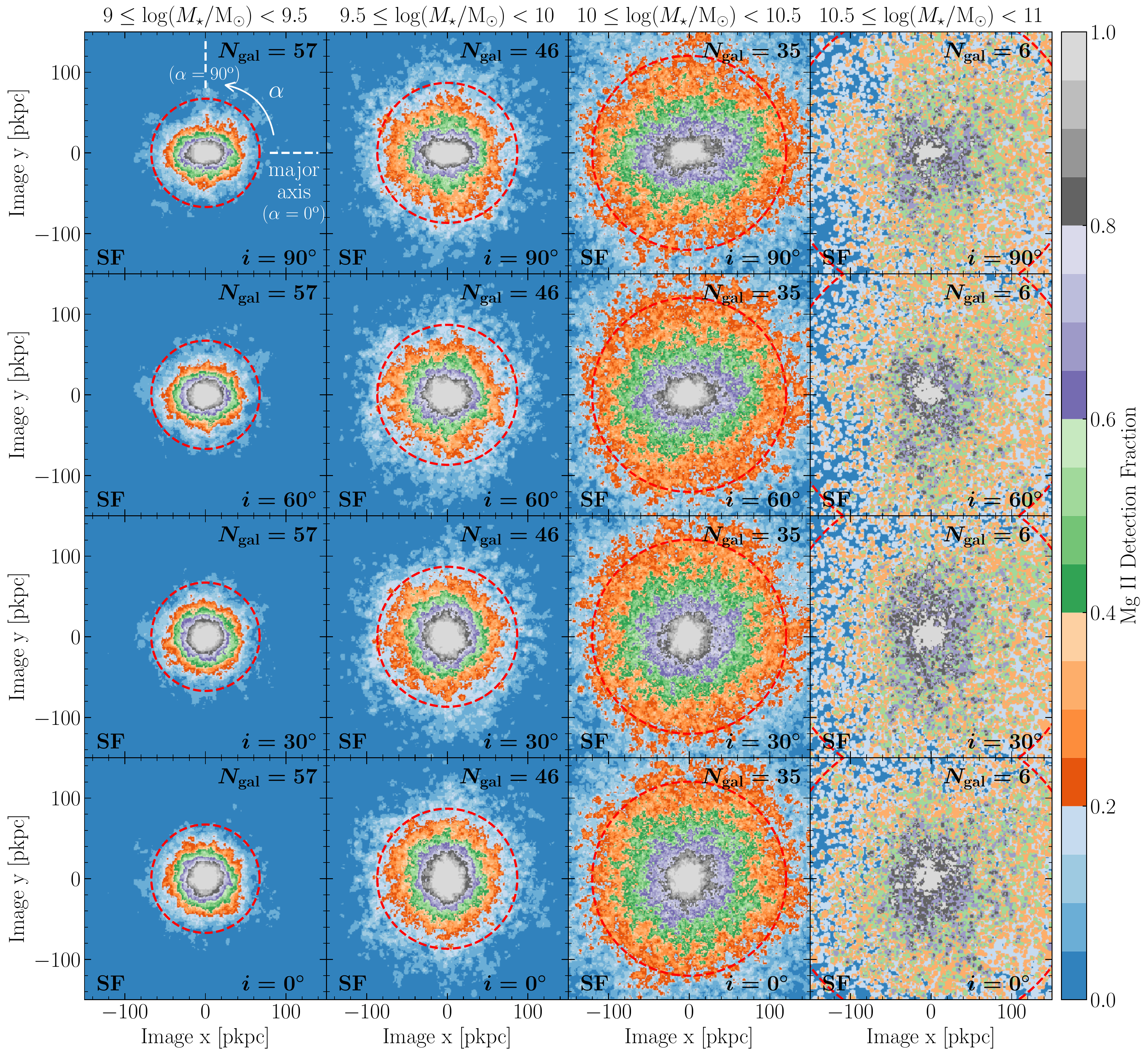}
    \caption{
        \mgII\ detection fraction of star-forming galaxies
        projected at different inclination angles
        and grouped into different stellar mass bins.
        The color maps represent 
        the \mgII\ detection fraction,
        which is calculated by dividing 
        for each pixel the 
        number of galaxies with \mgII\ gas ``detected''
        ($ N_\mathrm{MgII} \geq 10^{11.5}$\percmsq)
        by the total number of galaxies within the stack
        (labeled at the upper right of each panel).
        Similar to Figure~\ref{fig:mg2logN_sf_16p},
        each column shows the result for galaxies
        with different stellar masses,
        and each row represents galaxies projected
        at different inclination angles $i$ 
        before stacking.
        Each red dashed circle shows the 
        median 0.5\rvir\ of the galaxies in the stack.
        The \mgII\ gas is  
        not spherically distributed and extends to 
        larger radii for more massive galaxies.
        This analysis only includes the \mgII\ gas 
        within \rvir\ of individual galaxies.
        }
    \label{fig:mg2detfrac_sf_16p} 
\end{figure*}

We examine how the morphology and
radial extent of the \mgII\ gas
vary with galaxy stellar masses
and differ between star-forming and quiescent galaxies.
We project each galaxy at fixed inclination angles,
and we produce the 
corresponding column density maps 
of the \mgII\ gas within \rvir.
Then, we stack these individual maps based on
whether the galaxies
are star-forming or quiescent and their stellar masses.
For each pixel of each stack, 
we count the number of galaxies
with \mgII\ gas ``detected'',
i.e., $\log N_\mathrm{MgII} [\percmsq] \geq 11.5$,
and divide it by the total number of galaxies 
within the stack.
This generates a map of the ``\mgII\ detection fraction'',
which varies between 0 and 1 
if none or all of the galaxies
within the stack have ``detected'' \mgII\ gas. 
The maps in Figures~\ref{fig:mg2logN_sf_16p} 
and \ref{fig:mg2detfrac_sf_16p} 
show the median \mgII\ column density
and the \mgII\ detection fraction, respectively,
of all star-forming galaxies 
from $10^9$ \msununit\ to $10^{11}$ \msununit\ (columns)
and at inclination angles
of $i = $ 90\deg, 60\deg, 30\deg, and 0\deg\ (rows).

Figures~\ref{fig:mg2logN_sf_16p} and 
\ref{fig:mg2detfrac_sf_16p} demonstrate
two properties of the \mgII\ distribution around
star-forming galaxies.
First, the \mgII\ gas is not spherically distributed.  
The \mgII\ morphology changes with 
the projection angle of the galaxies;
for galaxies projected at 
higher (lower) inclination angles,
the column density maps show 
that the \mgII\ distribution 
is more flattened (isotropic), and
the contours of the \mgII\ detection fraction
are more elliptical (circular).
This is contrary to 
the circular contours expected 
on both sets of the 2D maps
at all projection angles
if the \mgII\ gas were spherically distributed.
The median column density maps in 
Figure~\ref{fig:mg2logN_sf_16p} 
largely resemble that of the galaxy example
in Figure~\ref{fig:pjdemo}
and show that the \mgII\ gas is
morphologically ``disky'' on average.
These average maps illustrate that 
the detectable \mgII\ gas possibly extends further away 
from the midplane than that of the example galaxy,
e.g., see the $10 \leq$ \logmstarmsun\ $<10.5$ panels.
In fact, the \mgII\ detection fraction maps 
demonstrate that even
at the $i=90$\deg\ projection (top row
of Figure~\ref{fig:mg2detfrac_sf_16p}), 
over 50\% of the galaxies (green contours) 
have detectable \mgII\ gas 
at least 20 pkpc above the midplane,
which is an order-of-magnitude thicker than gas disks
with a typical scaleheight of $\lesssim 2$ kpc 
\citep{vanderKruitFreeman2011,Kamphuis2013,
Zschaechner2015}.
Therefore, the shape of the contours 
of both the column density 
and detection fraction maps implies 
that generally neither a sphere nor a thin disk
describes the \mgII\ gas
(further discussed in Section~\ref{sec:discussion}).

\begin{figure}[thb]
    \centering
    \includegraphics[width=1.0\linewidth]{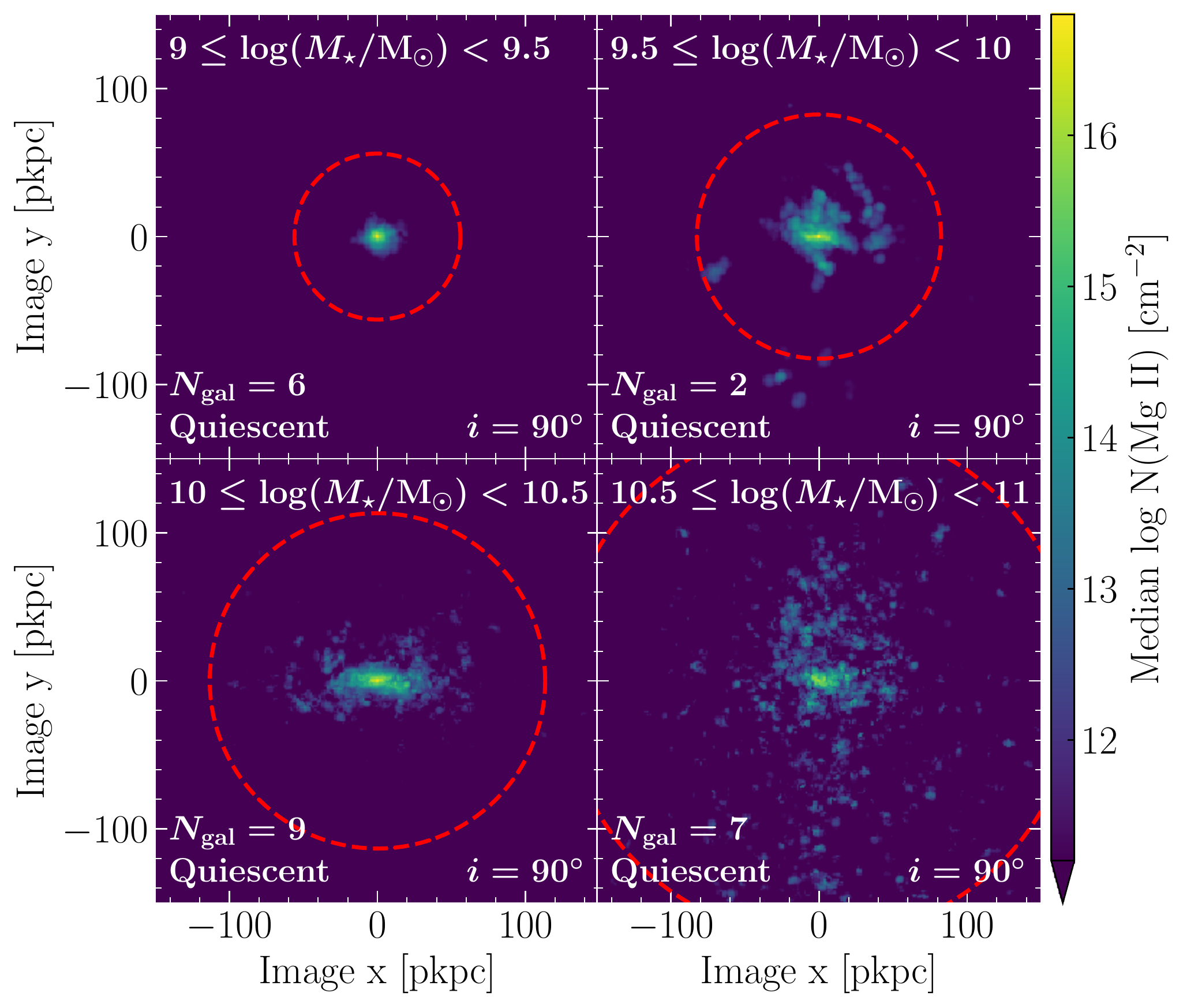}
    \caption{
        Median \mgII\ column density 
        around quiescent galaxies
        at $i=90$\deg\ projection.
        This plot is similar to the first row of 
        Figure~\ref{fig:mg2logN_sf_16p},
        but different panels show the stack of 
        quiescent galaxies with different stellar masses.
        The number of galaxies $N_\mathrm{gal}$
        in each stack
        is labeled at the lower left.  
        Each red dashed circle shows the 
        median 0.5\rvir\ of the galaxies in the stack.
        This analysis only includes the \mgII\ gas 
        within \rvir\ of individual galaxies.
        }
    \label{fig:mg2logN_qs_4p} 
\end{figure}

Secondly, the \mgII\ gas is more extended around 
star-forming galaxies with 
higher masses than around those with lower masses.
The detection fraction maps clearly 
demonstrate this trend.
For example, 50\% of the 
$10 \leq$ \logmstarmsun\ $< 10.5$ galaxies
(green contours)
``detect'' \mgII\ gas out to about 85 pkpc, 
but none of the 
$9 \leq $ \logmstarmsun\ $< 9.5$ galaxies
``detect''  \mgII\ gas at the same radius.  
The trend becomes less clear 
for the $10.5 \leq $ \logmstarmsun\ $< 11$ galaxies;
this is possibly due to small number statistics
with only six galaxies in the stack,
which also explain the less symmetric 
detection fraction contours 
compared to other mass bins.
Nonetheless, 
recall that we have only included the gas
within \rvir.  The trend is observed not only
in terms of the physical size (i.e., pkpc)
but also relative to the size of the virial halo.
The red dashed circle in each panel shows the 
median 0.5\rvir\ of the galaxies in the stack.
The \mgII\ gas around higher mass galaxies
still extends to larger radii relative to \rvir\ 
compared to lower mass galaxies
(also see Figure~\ref{fig:mg2detfrac_sf_rvir}
in the Appendix).

\begin{figure}[thb]
    \centering
    \includegraphics[width=1.0\linewidth]{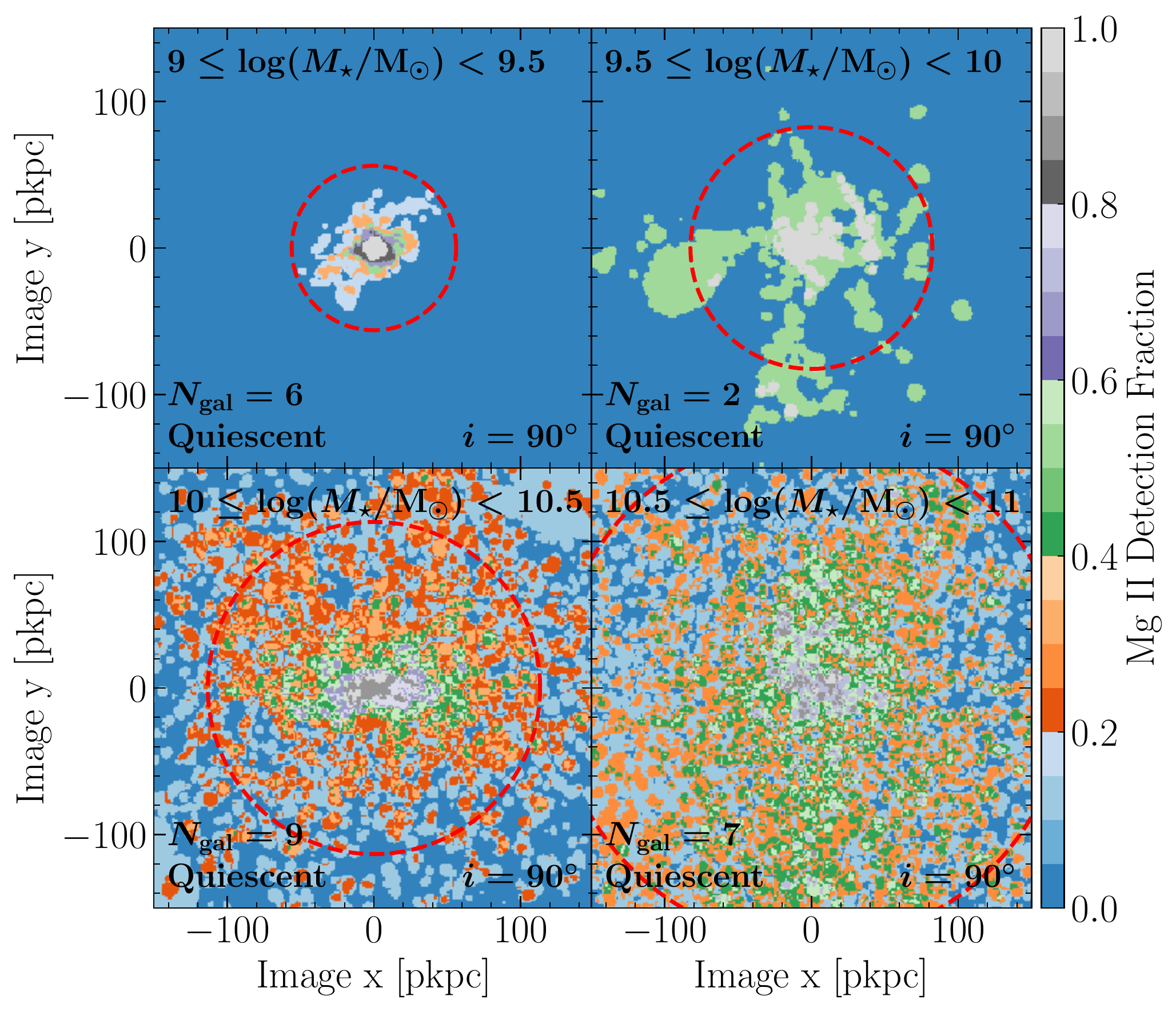}
    \caption{
        \mgII\ detection fraction around
        quiescent galaxies
        at $i=90$\deg\ projection.
        This plot is similar to the first row of 
        Figure~\ref{fig:mg2detfrac_sf_16p},
        but different panels show the stack of 
        quiescent galaxies with different stellar masses. 
        The number of galaxies $N_\mathrm{gal}$
        in each stack
        is labeled at the lower left.
        Each red dashed circle shows the 
        median 0.5\rvir\ of the galaxies in the stack.
        This analysis only includes the \mgII\ gas 
        within \rvir\ of individual galaxies.
        }
    \label{fig:mg2detfrac_qs_4p} 
\end{figure}

We repeat the same analysis for quiescent galaxies.
Figures~\ref{fig:mg2logN_qs_4p} and 
\ref{fig:mg2detfrac_qs_4p} show the 2D maps
of the median \mgII\ column density 
and the \mgII\ detection fraction, respectively, for
quiescent galaxies projected at $i=90$\deg.
We caution that each stellar mass bin only has 
a few quiescent galaxies (Table~\ref{tb:gal_mstar_count}),
so the results may be subject to 
small number statistics.  
Nevertheless, 
around quiescent galaxies with higher masses,
the \mgII\ gas has larger radial extent
in both physical size and relative to \rvir\ 
(also see Figure~\ref{fig:mg2detfrac_qs_rvir}
in the Appendix).
While Figures~\ref{fig:mg2logN_qs_4p} and 
\ref{fig:mg2detfrac_qs_4p} seem to suggest that
the \mgII\ distribution around quiescent galaxies 
is patchy,
the patchiness of the \mgII\ gas 
around individual galaxies will be washed 
out for a large number of galaxies,
creating a smooth distribution on average.
In fact, for each mass bin, 
if we randomly select the same number of star-forming
galaxies as quiescent galaxies,
then the maps for star-forming galaxies also show
patchier and less regular \mgII\ gas distributions
than Figures~\ref{fig:mg2logN_sf_16p} and
\ref{fig:mg2detfrac_sf_16p}.  
We also note that there exist 
$10 \leq$ \logmstarmsun\ $< 10.5$ quiescent galaxies
with \mgII\ gas preferentially residing 
near the midplane;
this pattern is also illustrated 
in the median column density map
(lower left panel of Figure~\ref{fig:mg2logN_qs_4p}).
But overall,
the \mgII\ gas around quiescent galaxies 
is potentially more isotropically distributed
compared to that around star-forming galaxies.


\begin{figure*}[thb]
    \centering
        \includegraphics[width=1.0\linewidth]{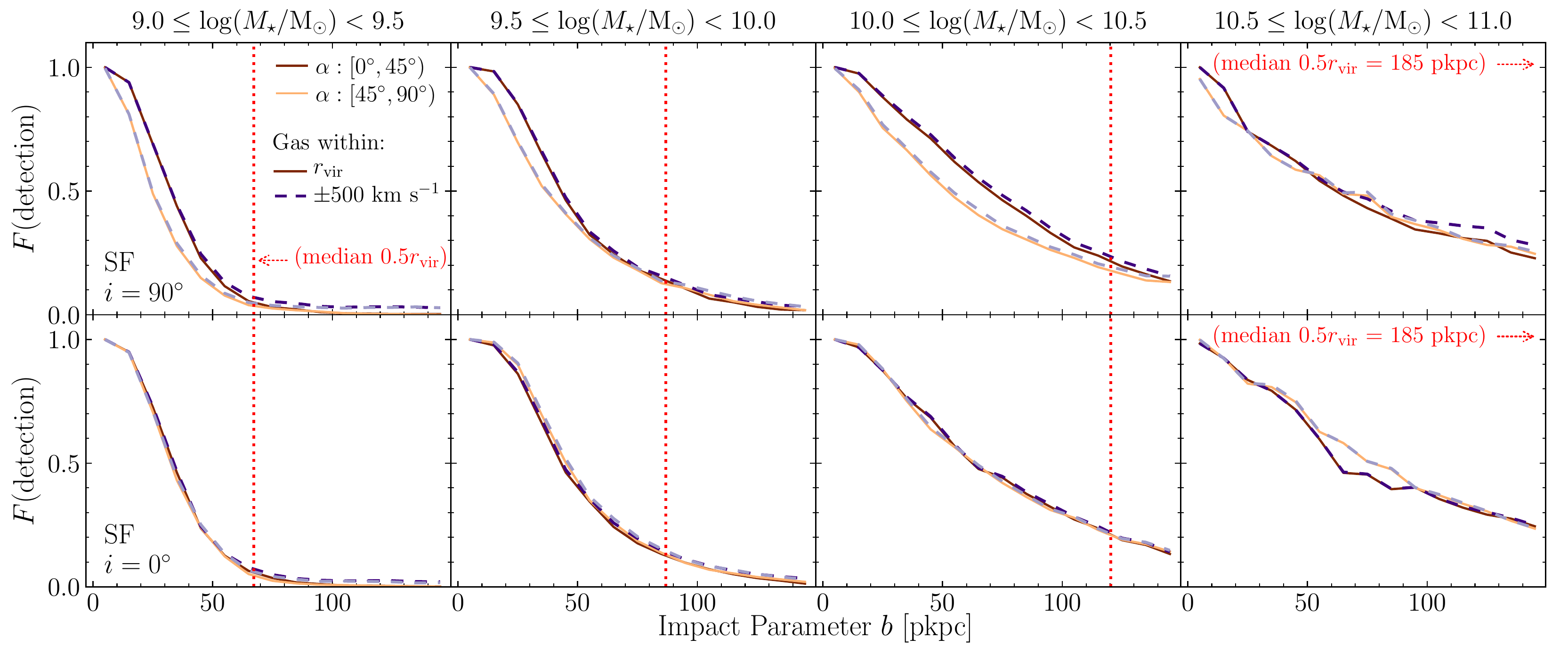}
    \caption{
        \mgII\ detection fraction 
        as a function of impact parameter $b$
        for star-forming galaxies projected at
        $i=90$\deg\ (i.e., seen edge-on, top row)
        and 0\deg\ (i.e., seen face-on, bottom row).
        Different columns show the results for 
        different stellar mass bins.  
        In each panel, 
        the solid and dashed curves
        represent the detection fractions obtained 
        from gas physically within \rvir\ 
        and within \deltavlos $= 500$\kms\
        from the galaxy systemic velocity,
        respectively.
        The darker (lighter) curve represents
        the azimuthal angle $\alpha$ range of 
        $0\deg \leq \alpha < 45\deg$ 
        ($45\deg \leq \alpha < 90\deg$), 
        i.e., closer to the galaxy major (minor) axes.
        The vertical red dotted line shows the 
        median 0.5\rvir\ of the galaxies in the stack.
        }
    \label{fig:mg2detfrac_sf_1d} 
\end{figure*}

While the \mgII\ detection fraction analysis
has been focusing on the gas within \rvir\
of individual galaxies,
observational studies typically associate 
\mgII\ gas with galaxy hosts
using a fixed LOS velocity window \deltavlos.
Because this potentially 
selects mis-assigned \mgII\ gas outside of \rvir\
(see Section~\ref{sec:mg2misassign}),
we explore how the mis-assigned \mgII\ gas
affects our results. 
Following Section~\ref{sec:mg2misassign},
we select the \mgII\ gas 
within \deltavlos $= 500$\kms\ 
from the galaxy systemic velocity
and repeat the detection fraction calculations.
As an illustration, 
Figure~\ref{fig:detfrac_sf_pm500} 
in the Appendix 
shows the new \mgII\ detection fraction maps
for $i=90$\deg\ star-forming galaxies.  
Comparing them with the original maps
obtained from the \mgII\ gas within \rvir\
(first row of  Figure~\ref{fig:mg2detfrac_sf_16p}),
the new light blue patches
near the edge of the maps ($\gtrsim 0.5$\rvir)
indicate that the detection fraction 
at these regions increases 
from around 0 to $\lesssim 0.1$.

Figure~\ref{fig:mg2detfrac_sf_1d} shows
the \mgII\ detection fractions for 
$i=90$\deg\ (top) and 0\deg\ (bottom) 
star-forming galaxies 
as a function of impact parameter.  
The solid and dashed curves represent 
the \mgII\ detection fraction calculated from 
the gas within \rvir\ and 
within the \deltavlos $= 500$\kms\ window
of individual galaxies, respectively.
We also bin the results by azimuthal angle $\alpha$,
which is the angle between the galaxy major axis
($x$-axis on the map) and the line joining the
center of the galaxy and each pixel;
see the illustration in the top left 
panel of Figure~\ref{fig:mg2detfrac_sf_16p}.
The darker (lighter) line
represents smaller (larger) azimuthal angles,
i.e., closer to the galaxy major (minor) axes.
First, 
regardless of how the gas is selected,
the \mgII\ detection fraction 
at a fixed impact parameter is higher (lower)
at smaller (larger) azimuthal angle 
when the galaxies are projected edge-on 
(i.e., $i=90$\deg).
Such difference largely disappears in 
face-on ($i=0$\deg) galaxy projections.
This reiterates the result that \mgII\ gas 
is not spherically distributed 
but preferentially resides near the midplane.
Second, for the same azimuthal angle bin,
selecting \mgII\ gas at 
\deltavlos\ $\leq 500$\kms\ (dashed)
produces a higher \mgII\ detection fraction than 
selecting \mgII\ gas physically within \rvir\
(solid).
The difference is around 0.02 in magnitude
and is more obvious 
at large impact parameters (e.g., $\gtrsim0.5$\rvir);
the 0.02 difference
represents an increase of over tens of percent 
at impact parameters where the \mgII\ within \rvir\ 
produces a low detection fraction.
Hence, 
our result implies that the mis-assigned \mgII\ gas 
will elevate the \mgII\ detection rate measured 
in random sightlines around galaxies.
As we will show in Section~\ref{ssec:rotation},
the mis-assigned gas has a more significant effect
on detecting corotating \mgII\ gas.

\subsection{Rotational Structure of the \mgII\ Gas}
\label{ssec:rotation}

Motivated by observational studies showing 
corotation between the \mgII\ gas and the galaxy disk,
we focus on the \mgII\ gas that corotates with 
the \EAGLE\ galaxies and 
examine the corotating gas structure.  
Similar to calculating the \mgII\ detection fraction, 
here we determine the 
fraction of \mgII\ that is corotating and detectable.
We project each galaxy at fixed inclination angles
and produce the 
\mgII-column-density-weighted LOS velocity maps.  
Because we orient the galaxies 
with the receding side at the $+x$-direction,
a $x > 0$ ($< 0$) pixel with a net redshift (blueshift)
indicates corotation.
We flag the pixels with \mgII\ gas 
``detected'' ($\log N_\mathrm{MgII} [\percmsq] \geq 11.5$)
and corotating.  
Then, at every pixel of each galaxy stack, 
we count the number of galaxies flagged
and divide it by total number of galaxies in the stack.  
The outcome measures how often we ``detect'' 
corotating \mgII\ gas among all galaxies.

\begin{figure*}[htb]
    \centering
    \includegraphics[width=0.95\linewidth]{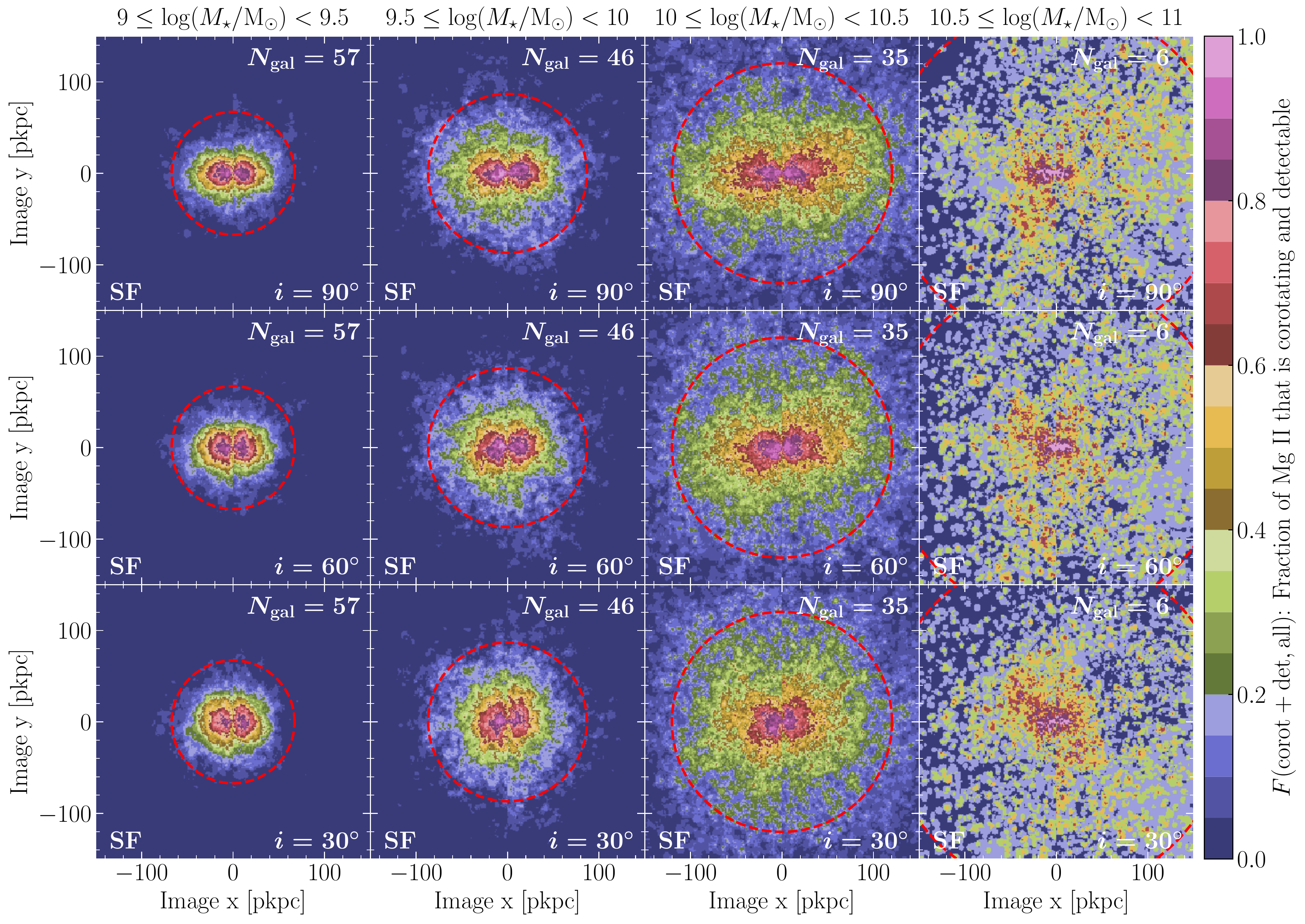}
    \caption{
        Fraction of \mgII\ that is 
        corotating and detectable
        for star-forming galaxies
        projected at different inclination angles
        and grouped into different stellar mass bins.
        The color maps represent 
        the fraction of \mgII\ that is 
        corotating and detectable,
        which is calculated by dividing the 
        number of galaxies with \mgII\ gas ``detected''
        ($ N_\mathrm{MgII} \geq 10^{11.5}$\percmsq)
        and corotating
        by the total number of galaxies in the stack
        (labaled at the upper right of each panel).
        Each column shows the result for galaxies
        in a different stellar mass bin
        (labeled at the top of each column).
        Each row represents galaxies projected at
        a different inclination angle $i$ 
        before stacking
        (labeled at the lower right of each panel).
        Each red dashed circle shows the 
        median 0.5\rvir\ of the galaxies in the stack.
        This analysis only includes the \mgII\ gas 
        within \rvir\ of individual galaxies.
        }
    \label{fig:mg2corot_sf_12p} 
\end{figure*}

Figure~\ref{fig:mg2corot_sf_12p} shows the
fraction of \mgII\ gas within \rvir\ 
of star-forming galaxies that is corotating
and detectable.
Different rows and columns show the results
for different galaxy inclination angles
and stellar mass bins, respectively.
We do not show the $i=0$\deg\ projection,
because the fraction of Mg II that is 
corotating and detectable becomes an ill-defined
quantity at $i=0$\deg;
the question of whether the detectable \mgII\ gas 
at individual $x > 0$ ($< 0$) pixel
shows a net redshift (blueshift) 
matches with that expected from an 
$i=0$\deg\ rotating disk is ill-defined,
because the latter produces zero Doppler shift,
i.e., neither blueshifted nor redshifted.
Comparing different columns 
of Figure~\ref{fig:mg2corot_sf_12p} 
shows that for more massive galaxies,
\mgII\ gas is more frequently 
detectable and corotating
at large projected radii compared to 
less massive galaxies.  
The trend becomes less clear 
for the highest mass bin 
of $10.5 \leq$ \logmstarmsun\ $< 11$,
mostly likely due to small number statistics 
with only a few galaxies in the stack.
Nevertheless, in general,
both the \mgII\ detection fraction
and the fraction of \mgII\ that is 
corotating and detectable
demonstrate the same trend that 
(corotating) \mgII\ gas is more extended 
around higher mass galaxies
(at least for galaxies with 
\logmstarmsun\ $< 10.5$).

\begin{figure*}[thb]
    \centering
    \includegraphics[width=1.0\linewidth]{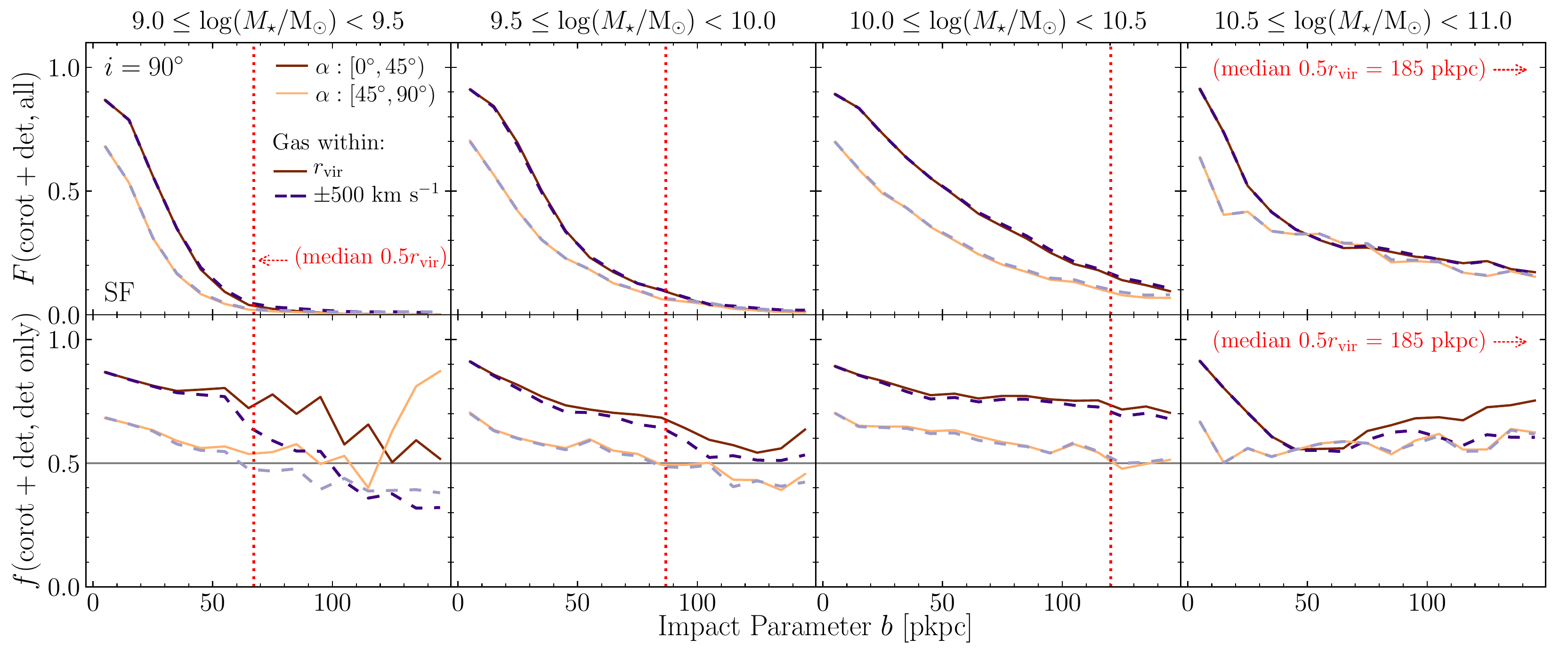}
    \caption{
        Fractions of corotating \mgII\ 
        as a function of impact parameter $b$
        for star-forming galaxies projected at
        $i=90$\deg\ (i.e., seen edge-on).  
        The top row shows
        the fraction of \mgII\ that is corotating
        and detectable calculated using all pixels
        in individual galaxy stacks,
        $F(\mathrm{corot+det}, \mathrm{all})$.
        The bottom row shows
        the fraction of detectable \mgII\ 
        that is corotating,
        $f(\mathrm{corot+det}, \mathrm{det\ only})$,
        which is calculated using only the 
        pixels with detectable \mgII.
        Different columns show the results for 
        different stellar mass bins.
        In each panel,
        the darker (lighter) line represents
        the azimuthal angle $\alpha$ range of 
        $0\deg \leq \alpha < 45\deg$ 
        ($45\deg \leq \alpha < 90\deg$),
        i.e., closer to the galaxy major (minor) axes.
        The solid and dashed curves
        represent the fractions obtained 
        from gas physically within \rvir\ 
        and within \deltavlos $= 500$\kms\
        from the galaxy systemic velocity,
        respectively.
        The vertical red dotted line 
        shows the median 0.5\rvir\ of the 
        galaxies in the stack.
        At a fixed impact parameter,
        both the top and bottom rows show that
        the fraction of corotating \mgII\ gas
        increases towards
        the galaxy major axis.
        The difference between selecting \mgII\ gas
        within \deltavlos\ and \rvir\ becomes
        prominent at impact parameters
        $\gtrsim0.25$\rvir.
        }
    \label{fig:mg2corotfrac_1d_sf} 
\end{figure*}

While the contour shape of 
the \mgII\ corotation maps in 
Figure~\ref{fig:mg2corot_sf_12p}
also changes with the projected inclination angles, 
the contours take a different shape from 
those of the detection fraction.
Especially near the galaxy center and regions
with high fractions of 
corotating and detectable \mgII\ (e.g., $\geq50$\%),
the contours
resemble a dumbbell shape
with the two lobes lying along 
the galaxy major axis (i.e., $x$-axis on the map).  
This implies that the fraction of \mgII\ 
that is corotating and 
detectable is reduced near the galaxy minor axis;
we will discuss possible explanations 
(e.g., outflows) in Section~\ref{ssec:structure_sf}.

Instead of showing the 2D maps, 
the top row of Figure~\ref{fig:mg2corotfrac_1d_sf} 
shows the fraction of \mgII\ that is corotating and 
detectable as a function of impact parameter
for star-forming galaxies at the $i=90$\deg\ projection.  
Similar to Figure~\ref{fig:mg2detfrac_sf_1d},
the darker (lighter) curve
represents smaller (larger) azimuthal angle $\alpha$,
i.e., closer to the galaxy major (minor) axes.
The solid and dashed curves show the results
obtained from the \mgII\ gas within \rvir\
and within \deltavlos $= 500$\kms\ 
from the systemic velocity of individual galaxies,
respectively.
Regardless of which \mgII\ selection method we use,
clearly for all mass bins (shown in separate panels), 
the fraction decreases towards 
the galaxy minor axis for a fixed impact parameter.
This indicates a paucity of 
net corotating \mgII\ gas towards the minor axis.

While the fraction of \mgII\ that is
corotating and detectable decreases sharply
with impact parameter
and does not seem to depend on 
how we select the \mgII\ gas,
we emphasize that this steep decline
largely results from the sharp drop
in the detection fraction.
In other words, it does not
necessarily imply a transition
of the \mgII\ gas from having a net corotation
to a lack thereof.
Our calculation is analogous to examining how often
we ``detect''  corotating \mgII\ gas
for random sightlines around galaxies.
But from the observers' perspective,
the more interesting question is
how often they measure a net corotation
in sightlines that detected \mgII\ gas
(e.g., \citealt{Martin2019},
also see the Ly$\alpha$ ``corotation fraction'' analysis 
in \citealt{FrenchWakker2020}).
Therefore,
we modify our calculation to answer this question.
Instead of dividing the number of galaxies
with corotating \mgII\ ``detected''
by the total number of galaxies in the stack,
we divide it by the number of galaxies
with ``detected'' \mgII\ gas.
The outcome represents the 
fraction of detectable \mgII\ gas that is 
corotating.

\begin{figure}[thb]
    \centering
    \includegraphics[width=1.0\linewidth]{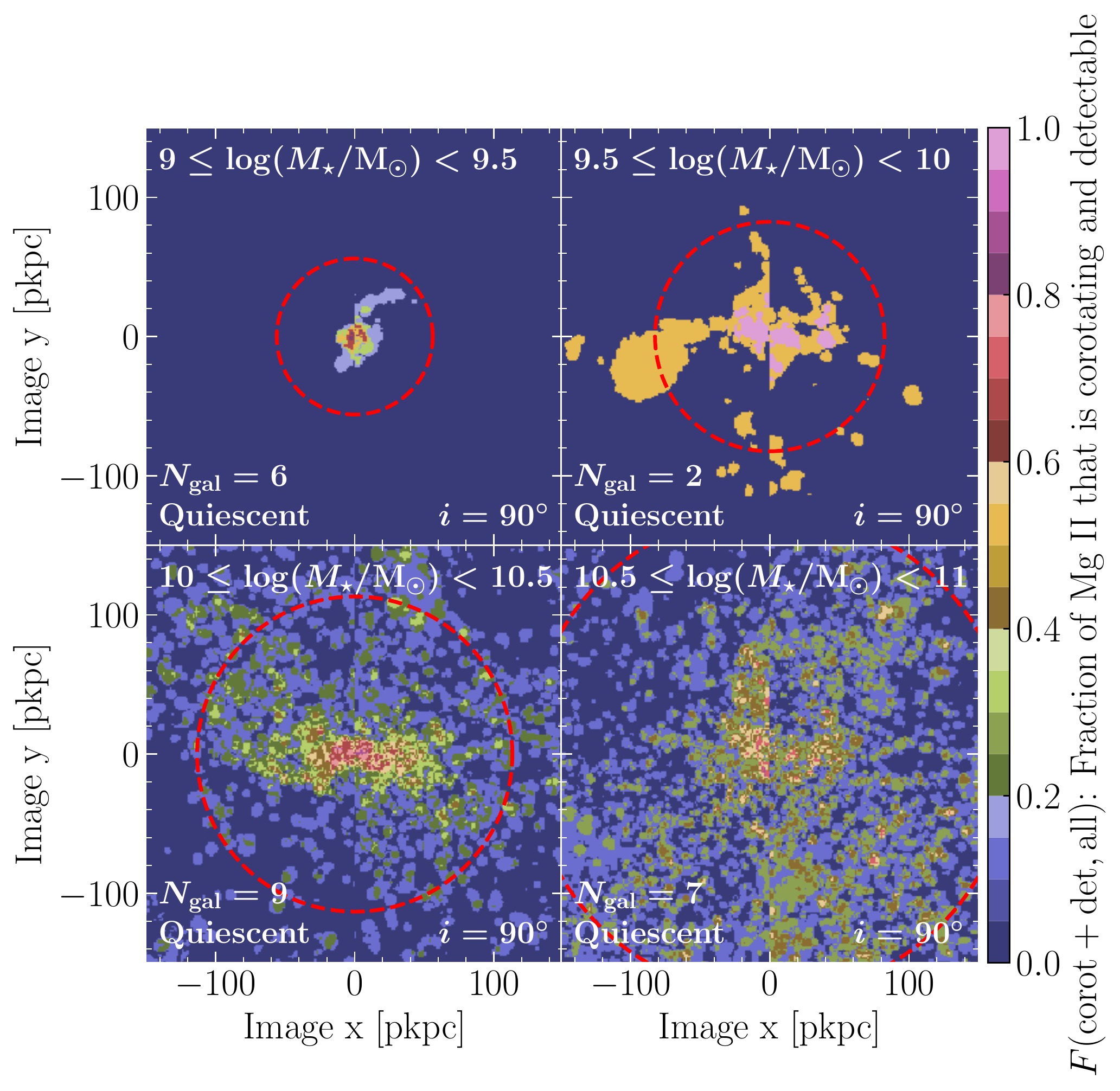}
    \caption{
        Fraction of \mgII\ that is 
        corotating and detectable
        for quiescent galaxies
        at $i=90$\deg\ projection.
        Each panel shows the galaxy stack of 
        different mass bins.
        The number of galaxies $N_\mathrm{gal}$
        in each stack is labeled at the lower left.
        Each red dashed circle shows the 
        median 0.5\rvir\ of the galaxies in the stack.
        This analysis only includes the \mgII\ gas 
        within \rvir\ of individual galaxies.
        }
    \label{fig:mg2corot_qs_4p} 
\end{figure}

The bottom row of Figure~\ref{fig:mg2corotfrac_1d_sf}
shows the fraction of detectable \mgII\ gas that is 
corotating as a function of impact parameter
for $i=90$\deg\ star-forming galaxies.
Clearly, the results depend on
whether we select the gas that is 
within \rvir\ (solid) or 
within the \deltavlos\ $=500$\kms\ window (dashed).
While the results from 
both \mgII\ selections share some characteristics,
such as exhibiting an azimuthal dependence
and declining slightly with impact parameter,
the difference between
using the two selection methods
becomes prominent at impact parameters beyond
$\approx50$ pkpc or 0.25\rvir.
Compared to selecting the \mgII\ gas within \rvir,
the \mgII\ gas selected by \deltavlos\ $\leq 500$\kms\
has a lower fraction of detectable \mgII\ gas that is 
corotating.  
This means that the \mgII\ outside of \rvir\
reduces the frequency of detecting
net corotating \mgII\ gas
at a fixed impact parameter.
We will discuss how this impacts
the observational analysis of corotating \mgII\ gas in
Section~\ref{ssec:obs_corot_rvir}.
We caution that
especially in the lowest mass bin,
although the fraction 
obtained from gas within \rvir\ (solid)
shows a slight increase 
at large impact parameters ($\gtrsim130$ pkpc),
this is due to the rare \mgII\ detection 
(Figure~\ref{fig:mg2detfrac_sf_1d}),
which makes the measured fraction 
of detectable \mgII\ that is corotating 
very noisy.


As for the quiescent galaxies,
it is less clear whether or not
their \mgII\ gas is generally rotating
as is the case for
the star-forming galaxy counterparts.
Figure~\ref{fig:mg2corot_qs_4p} shows 
the fraction of \mgII\ that is corotating and 
detectable for
quiescent galaxies projected at $i=90$\deg\ 
(analogous to the top row of 
Figure~\ref{fig:mg2corot_sf_12p}),
and the contours do not show a particular pattern.
This could again be a result of 
small number statistics, 
which we have already seen in 
the detection fraction maps.
But still, there exist 
quiescent galaxies of $10 \leq$ \logmstarmsun\ $< 10.5$
having \mgII\ gas that is 
morphologically and kinematically ``disky''.  
This can be seen marginally in 
Figure~\ref{fig:mg2corot_qs_4p},
but the contours are too irregular to 
make a general description.
Hence, 
the \mgII\ detection fraction and 
the fraction of \mgII\ that is corotating and 
detectable potentially suggest that
quiescent and star-forming galaxies 
have different \mgII\  morphology 
and kinematic structure,
but the poor statistics for quiescent galaxies
make this result inconclusive.

\section{Discussion}
\label{sec:discussion}

We have shown that how the 
detection fraction of \mgII\ and the
identification of corotating \mgII\ gas vary
with galaxy properties.
We have also raised the concern that
the \mgII\ gas selected by a LOS velocity window
(\deltavlos\ $\leq 500$\kms)
but physically outside of \rvir,
i.e., the mis-assigned \mgII\ gas, 
possibly affects the observational analysis of the CGM.
In this section,
we will discuss how 
this mis-assigned \mgII\ gas affects 
the observational analysis
of  corotating gas in sightline studies.
We will also interpret our results regarding  
the morphological and kinematic structure 
of the \mgII\ gas
and discuss recent related work in 
observations and simulations.

\subsection{How does the mis-assigned \mgII\ gas 
affect the \mgII\ detection and corotation analysis
in observational studies?}
\label{ssec:obs_corot_rvir}

Quasar sightline studies have established 
the steep decline in the covering factor 
(i.e., detection rate)
and the strength of the \mgII\ absorption systems  
with impact parameter
\citep{Chen2010,Nielsen2013_ii,Lan2014,LanMo2018,Huang2020}.
Our \mgII\ detection fraction analysis
agrees with this result.  
However, we also demonstrated
that if we select the \mgII\ gas using the 
\deltavlos $= 500$\kms\ window instead of 
requiring the gas to be physically within \rvir,
then the detection fraction 
increases by around 0.02 in magnitude.
This can correspond to an increase
of several tens of percent 
for lower mass galaxies and 
at large impact parameters 
(e.g., $\gtrsim100$ pkpc),
where the gas within \rvir\ 
only gives a low \mgII\ detection fraction
of the order of 0.01
(Figure~\ref{fig:mg2detfrac_sf_1d}).
As a result, the increase in detection fraction
at large radii
makes the circumgalactic \mgII\ gas seem
more extended around galaxies than it is.
Because observers often use \deltavlos\ 
to identify the \mgII\ gas around target galaxies, 
the mis-assigned \mgII\ gas outside of \rvir\
will increase the number of \mgII\ systems detected
and/or the strength of \mgII\ systems measured
at large radii.
This potentially affects the 
anticorrelation between 
covering fraction (and \mgII\ strength)
and impact parameter derived 
from quasar sightline observations,
such as increasing the uncertainties of 
the fit between the two quantities
or weakening the anticorrelation.

The mis-assigned \mgII\ gas outside of \rvir\
but within the \deltavlos\ window
has a significant effect on 
identifying corotating \mgII\ gas.  
The bottom row of Figure~\ref{fig:mg2corotfrac_1d_sf}
shows the fraction of detectable \mgII\ gas 
that is corotating as a function of impact parameter.
From the observers' perspective, 
this fraction represents how likely it is to 
find corotating \mgII\ gas in \mgII\ detected 
sightlines.
If sightlines intersect randomly moving gas,
then there should be an equal number 
of sightlines intersecting 
corotating and non-corotating gas.
Hence, at a fixed impact parameter,
observers expect to detect 
the fraction exceeding
0.5 if the \mgII\ gas generally 
has a net corotation (grey horizontal lines). 
At impact parameters beyond
$\approx50$ pkpc or 0.25\rvir,
the plots show that the fraction is lower when 
the \mgII\ gas is selected by \deltavlos\ (dashed)
instead of within \rvir\ (solid).
For example,
near the major axis ($\alpha < 45$\deg)
of the $9.0 \leq $ \logmstarmsun\ $< 9.5$ 
star-forming galaxies,
the fraction drops to 0.5 at 100 pkpc
if we select the \mgII\ gas by \deltavlos,
but the detectable \mgII\ within \rvir\ 
actually shows net corotation out to
130 pkpc ($\approx$ \rvir).
Observers typically identify the \mgII\ gas
around target galaxies using the \deltavlos\ window.
Therefore, our result implies that
at large impact parameters
where \mgII\ gas is less often detected,
observers will measure a lower detection rate of 
\mgII\ corotating gas among 
the \mgII\ detected sightlines, because 
the mis-assigned \mgII\ gas contaminates the signal.
This leads observers to 
underestimate the spatial extent
of \mgII\ corotating gas,
i.e., the measured fraction of 
detectable \mgII\ that is corotating drops below 0.5 
at too small of an impact parameter.

\subsection{Coherent \mgII\ Gas Structure Around Star-forming Galaxies}
\label{ssec:structure_sf}

While we have illustrated that
\mgII\ gas disks exist (e.g., Figure~\ref{fig:pjdemo}),
the median \mgII\ column density maps
(Figure~\ref{fig:mg2logN_sf_16p})
and detection fraction maps 
(Figure~\ref{fig:mg2detfrac_sf_16p})
indicate that a thin disk
does not describe the typical \mgII\ distribution
around star-forming galaxies.  
The \mgII\ gas clearly does not have 
a spherical distribution either,
because the contours are non-circular 
with their shape changing with 
the galaxy inclination projection angle.
Instead, the maps show that
the \mgII\ gas distribution is axisymmetric.
Furthermore, in regions 
near the projected galaxy major axes
(galaxy orientation defined by \jstar),
the corotating \mgII\ gas is 
more frequently detected.
The contours of the \mgII\ corotation maps show 
two lobes aligning with the galaxy major axes 
(Figure~\ref{fig:mg2corot_sf_12p}),
and the 1D profiles also show that 
at a fixed impact parameter, 
the fraction of detectable \mgII\ 
that is corotating decreases with increasing 
azimuthal angle 
(bottom row of Figure~\ref{fig:mg2corotfrac_1d_sf}).
Altogether, 
these results not only show that 
the \mgII\ gas is not spherically distributed
and has an axisymmetric structure,
but that the axis of symmetry aligns with that 
of the rotation of the stars and the \mgII\ gas. 
In fact, the 50\% contours 
of the \mgII\ detection fraction suggest that 
for half of the galaxies,
these rotating \mgII\ structures 
possibly reach over 20 pkpc from the midplane.
An axisymmetric, rotating \mgII\ structure 
can also explain the low fraction 
of corotating and detectable \mgII\ gas
at large azimuthal angles, 
i.e., near the projected galaxy minor axes.
Because the projected velocity from the 
tangential velocity component 
is small near the minor axis,
any turbulence overwhelms the net corotation signature
and decreases the rate of detecting the corotating gas.

\begin{figure}[tbh]
    \centering
    \includegraphics[width=1.0\linewidth]{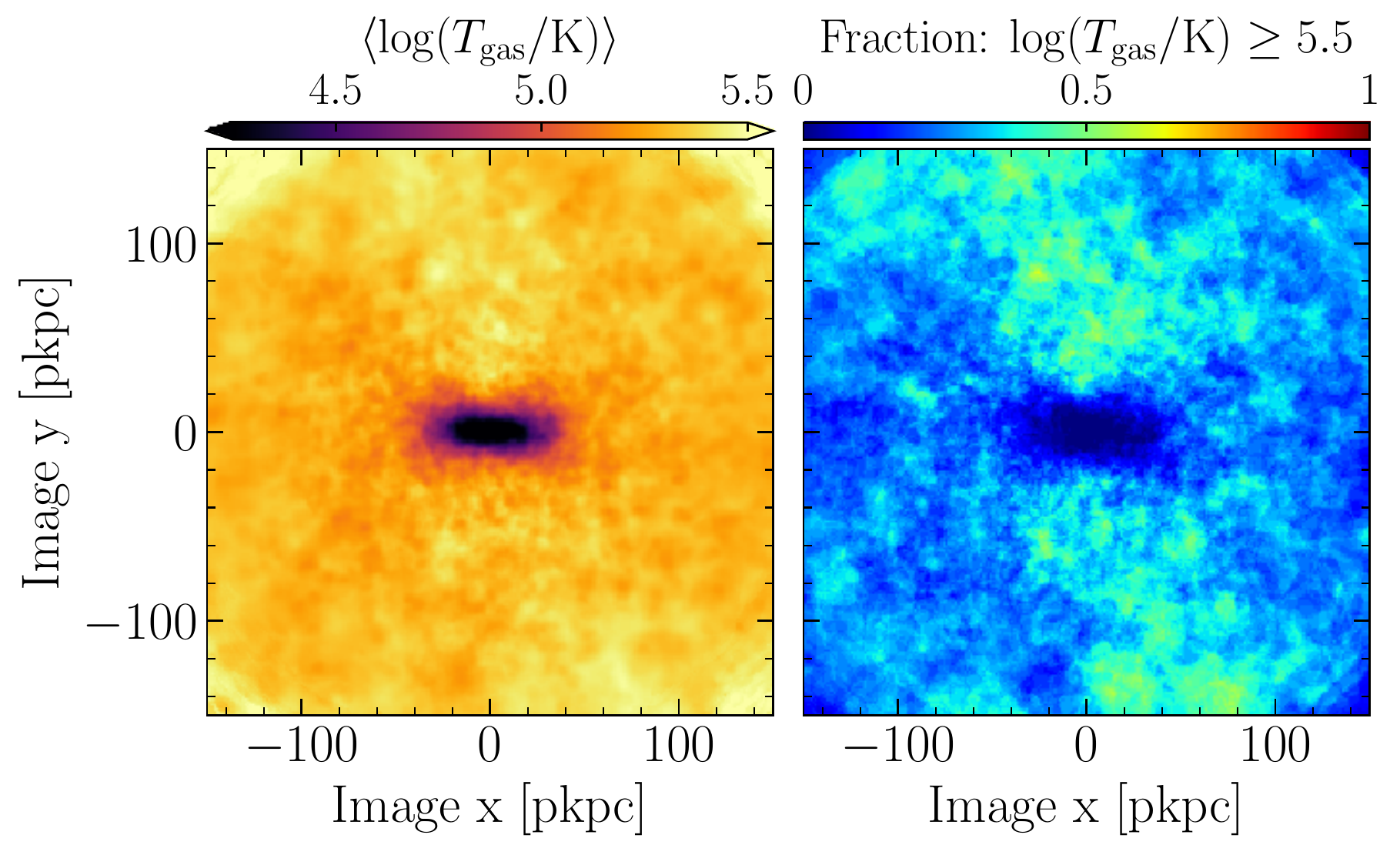}
    \caption{
        Average temperature of gas within \rvir\
        of the $9.5 \leq $ \logmstarmsun\ $< 10.0$ 
        star-forming galaxies projected at $i=90$\deg.
        \textit{(left)} Mean gas temperature
        $\langle \log (T_\mathrm{gas}/\mathrm{K})\rangle$ 
        of the stacked galaxies.
        \textit{(right)} Fraction of galaxies
        with $\log (T_\mathrm{gas}/\mathrm{K}) \geq 5.5$.
        Both panels illustrate that 
        gas near the projected galaxy minor axes
        typically has a higher temperature.
        }
    \label{fig:logT_mstar_example}
\end{figure}

A possible source of turbulence is 
a wind blown out perpendicular to the disk plane
\citep{DeYoungHeckman1994}
that kinematically disturbs the CGM
\citep{Heckman2017,Martin2019}.
Around the \EAGLE\ galaxies,
we find the signature of hot winds
from the higher gas temperature 
in the biconical regions above and below the disk plane.
As an illustration, 
the left panel of Figure~\ref{fig:logT_mstar_example}
shows the higher mean gas temperature 
($\langle\log T_\mathrm{gas}\rangle$)
near the minor axes 
of the $9.5 \leq $ \logmstarmsun\ $< 10.0$
star-forming galaxies.
Similarly, the right panel shows
a higher fraction of galaxies
with average gas temperature 
$\geq 10^{5.5}$ K (cyan) near the minor axes.  
The higher temperature near the minor axis 
also implies that the wind signature can be observed in 
the warm-hot and/or hot phases traced by higher ions,
but this is beyond the scope of this paper.
\citet{Mitchell2020outflow} have also recently
shown that the CGM of \EAGLE\ galaxies
typically exhibits a bipolar outflow pattern
aligning with the galaxy minor axis
(see their Figure~7).

It is worth noting that
theoretical work and recent simulations have raised 
the concern of how the cool gas (e.g., traced by \mgII)
survived entrainment by the hot winds
\citep{Schneider2018,GronkeOh2020}.
Tackling this question requires 
simulations with parsec-scale resolution
in the low-density CGM
and has presented challenges 
for large-scale cosmological simulations.
For example, 
both the high-resolution 25 Mpc \EAGLE\ volume
and \texttt{TNG50} \citep[][Figure 1]{Nelson2020} 
have an average comoving particle separation
or gas resolution of about 
1 kpc at densities of 
$n_\mathrm{H}\sim 10^{-2}$ \percmcube,
and the resolution worsens (improves)
at lower (higher) density regions.
Because most cosmological simulations
focus the computational resources
at denser regions and coarsely resolve the CGM,
recent efforts applied new refinement schemes
to enhance the CGM resolution
in cosmological zoom simulations 
of individual objects.  
For example, 
\citet{vandeVoort2019} used the \AREPO\ 
moving-mesh hydrodynamics code \citep{Springel2010} 
with an additional uniform spatial refinement
to force a minimum cell size of 1 pkpc
in their Milky Way mass galaxy simulations,
whereas \citet{Hummels2019}
and the FOGGIE simulations \citep{Peeples2019}
used the adaptive mesh refinement code 
\ENZO\ \citep{Bryan2014}
with their independently developed refinement techniques
and resolved spatial scales of about 500 comoving-pc 
out to 100 comoving-kpc in galactocentric radius.
However, for the higher gas densities 
typical of low-ionization absorbers,  
the gains in resolution are modest 
relative to the simulation analyzed here. 
Future work achieving  
even higher resolution in dense gas 
will be important to shed further light 
on the survival of \mgII\ clouds.


Our picture of the axisymmetric, rotating 
\mgII\ circumgalactic gas 
around star-forming galaxies
broadly agrees with 
recent results of CGM analyses 
using different cosmological simulations,
all of which establish a picture of the 
rotating CGM with significant angular momentum.
Using zoom-in simulations 
with the \EAGLE\ model,
\citet{Huscher2020} recently found that
the angular momentum vectors between
the hot and cold components of the CGM
are well-aligned and better than that of the
stellar disk at $z=0$.
They showed that
the cold gas has a higher specific
angular momentum than the hot gas.
The tangential velocities of the cold gas (and metal)
suggested that
the cold gas is primarily rotationally supported
out to 40 kpc in radius.
The hot gas has a lower tangential velocity
but still shows net rotation
out to the radius of 50 kpc,
implying that the hot CGM
is poorly described by hydrostatic equilibrium
\citep{Oppenheimer2018hse}.
The recent work of \citet{DeFelippis2020}
also showed that 
the cold CGM has 
a higher specific angular momentum
for $z\lesssim2$ galaxies in \illustrisTNG.
For their high-\jstarscalar\ (low-\jstarscalar)
galaxy subsample,
the angular momentum vector alignment
between the stellar component and the CGM
is stronger than (comparable to)
that found in \citet{Huscher2020}.
In addition,
they showed that winds and fountain gas
dominated the biconical polar region,
whereas the cold, high angular momentum gas
occupied a wedge near the planar region
on their single-quadrant 2D map.
This led to their conclusion of
a cylindrically symmetric CGM distribution.
We note that their
wedge is analogous to our 
two lobes along the galaxy major axis shown
on the \mgII\ corotation maps 
(Figure~\ref{fig:mg2corot_sf_12p}).
Earlier work by \citet{Kauffmann2016,Kauffmann2019}
also found a rotating CGM around Milky Way-like galaxies
in both \illustris\ and \illustrisTNG.
The CGM in \illustris\ rotates coherently
over 70 kpc
and has a larger vertical coherent length
than that in \illustrisTNG,
and the authors attributed the difference
to the change in the feedback prescriptions.


The rotating CGM in cosmological simulations
and our description of the
axisymmetric, rotating \mgII\ gas structure 
support the interpretation and kinematic modeling
from circumgalactic absorption measurements
and share similarities with 
\hI\ observations of nearby galaxies.
Quasar sightline observations 
found \mgII\ gas corotating with the galaxy disk,
but the \mgII\ gas spans broader velocity range
than a thin rotating disk can explain
\citep[][etc.]{Steidel2002,
Kacprzak2010,Kacprzak2011ApJ,Martin2019}.
Modeling the \mgII\ gas as a thick disk 
(with or without a rotation lag)
with a height of over 20 kpc 
\citep{Steidel2002,Kacprzak2010,Kacprzak2011ApJ}
or combining it with a radial inflow component
can plausibly reproduce 
the measured \mgII\ kinematics
\citep{Ho2017,HoMartin2020}.  
On the one hand, these models 
defy the general perception of a disk.
Stellar disks and gas disks typically have 
scaleheights below hundreds of parsec \citep{deGrijs1998}
and $\lesssim2$ kpc 
\citep[e.g.,][]{vanderKruitFreeman2011,Gentile2013},
respectivety, both of which are at least 
an order-of-magnitude smaller than the height 
of the modeled thick disk.
On the other hand,
\hI\ observations of nearby edge-on galaxies
found extra-planar \hI\ gas
several kpc or even $\approx$ 20 kpc
from the disk midplane,
e.g., NGC 891 \citep{Oosterloo2007}.
These \hI\ observations detect
a lag in rotation speed as 
the distance from the disk plane increases,
and some also measure
a decrease in
this vertical velocity gradient at the outer radii
\citep{Oosterloo2007,Zschaechner2012}.
Multi-component models were developed to reproduce the 
measured \hI\ column density distribution 
in different velocity bands, 
i.e., the \hI\ channel maps.
These models include some combinations of 
a thick disk with rotation lag,
radial flow, flare, warps,
and allow asymmetry between 
the approaching and receding sides of the rotation
\citep{Oosterloo2007,Zschaechner2012,Zschaechner2015,Kamphuis2013}.
Is it possible that the \mgII\ gas structure 
resembles that of the \hI\ gas 
but scaled up in size?
Future work can explore this question by 
focusing on individual galaxies 
in zoom-in cosmological simulations
and {creating models to analyze 
the \mgII\ channel maps the same way 
as the \hI\ channel maps in observational analyses.


\subsection{\mgII\ Gas Distribution Around Quiescent Galaxies}
\label{ssec:structure_quiescent}

The maps of the \mgII\ detection fraction
(Figure~\ref{fig:mg2detfrac_qs_4p})
and the distribution of detectable corotating \mgII\ gas
(Figure~\ref{fig:mg2corot_qs_4p})
potentially suggest that 
average quiescent galaxies have a more isotropic
\mgII\ distribution and less ``disky''
morphologically and kinematically  
compared to their
star-forming galaxy counterparts.
While our results for quiescent galaxies 
may be subject to 
small number statistics 
(Table~\ref{tb:gal_mstar_count} and 
Section~\ref{ssec:morphology}), 
the result that star-forming and quiescent galaxies
have different \mgII\ gas properties
is well supported by observational studies.
For example,
quiescent galaxies have a lower 
\mgII\ covering fraction than star-forming galaxies
\citep{Lan2014,Huang2020},
and the covering fraction drops further for 
massive, luminous red galaxies 
\citep[LRGs,][]{Huang2016,Chen2018}.
Comparison of our \mgII\ detection maps 
between star-forming and quiescent galaxies
agrees with this description,
and plotting the detection fraction
against impact parameter clearly 
demonstrates the lower detection rate 
for quiescent galaxies at a fixed impact parameter
(Figure~\ref{fig:mg2detfrac_qs_1d};
thick curves).
Also, strong \mgII\ systems are perferentially 
observed around star-forming galaxies,
and the strength of the \mgII\ absorption
shows an azimuthal dependence 
around star-forming galaxies but 
not around quiescent galaxies;
the latter led to the conclusion that \mgII\ gas 
around quiescent galaxies is
isotropically distributed \citep{Bordoloi2011,Lan2014}.
These differences in the CGM 
between star-forming and quiescent galaxies are not
limited to the low-ionization-state \mgII\ gas
but also apply to the higher ions.
For example, the highly ionized \oVIion\ ion
ubiquitously observed around 
$\sim$$L^{*}$ star-forming galaxies
is rarely detected around quiescent galaxies
in the COS-Halos survey \citep{Tumlinson2011}.
While this dichotomy 
is reproduced by both the 
\EAGLE\ \citep{Oppenheimer2016} 
and \illustrisTNG\ simulations \citep{Nelson2018},
Oppenheimer \etal\ showed that 
the observed dichotomy
largely reflects the higher halo mass of 
the quiescent galaxies compared to the
star-forming galaxies in the COS-Halos sample,
for which the \oVI\ fraction peaks at the 
halo virial temperature of 
the $\sim$$L^{*}$ star-forming galaxies.  
Nevertheless, both \EAGLE\ and \illustrisTNG\
predict that at a fixed halo mass,
the CGM gas mass fraction strongly correlates 
with the galaxy sSFR \citep{Davies2020}.
Hence, both observational 
and simulation studies have verified that 
star-forming and quiescent galaxies
have different circumgalactic gas properties.

\begin{figure*}[hbt]
    \centering
    \includegraphics[width=1.0\linewidth]{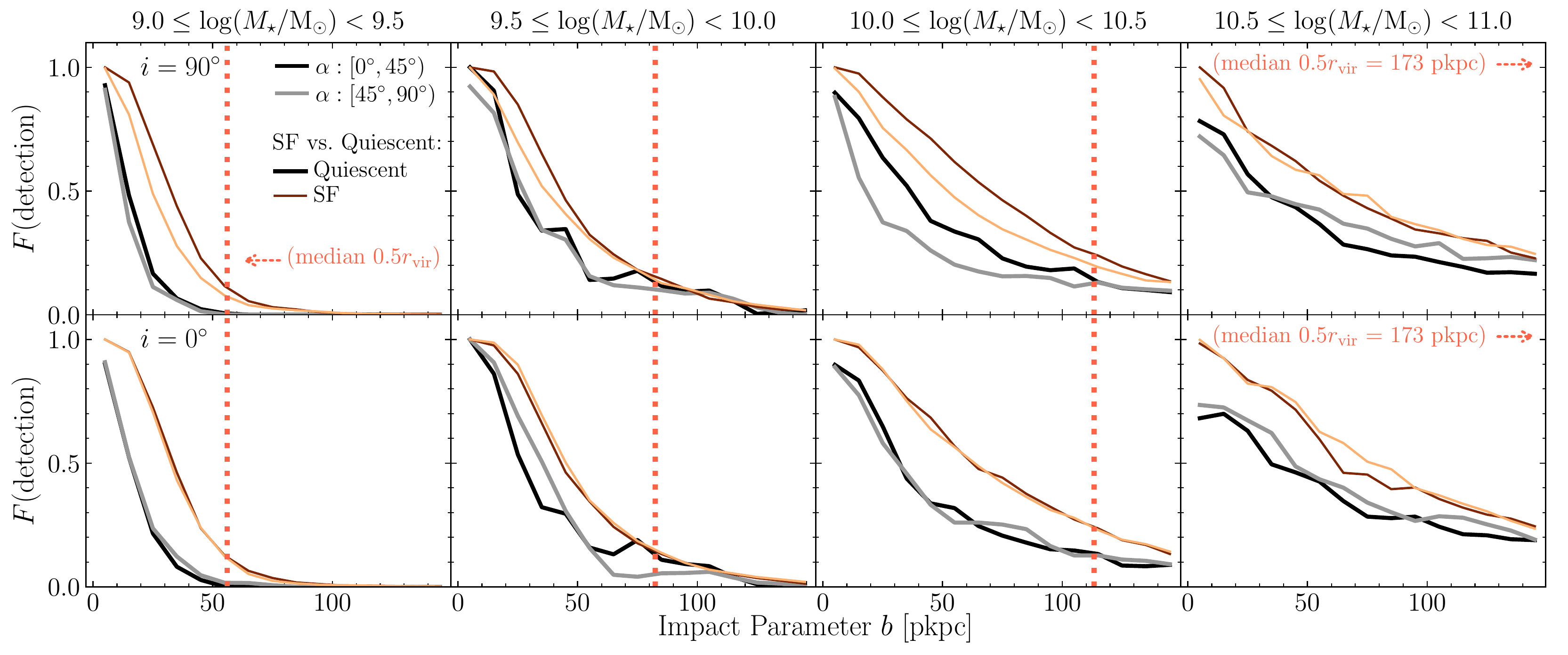}
    \caption{
        \mgII\ detection fraction 
        as a function of impact parameter $b$
        for quiescent and star-forming
        galaxies projected at
        $i=90$\deg\ (i.e., seen edge-on, top row)
        and 0\deg\ (i.e., seen face-on, bottom row).
        The solid curves
        represent the detection fractions obtained 
        from gas physically within \rvir\
        of quiescent (thick, greyish) and 
        star-forming (thin, reddish) galaxies.
        Similar to Figure~\ref{fig:mg2detfrac_sf_1d},
        different columns show the results for 
        different stellar mass bins,
        and the darker (lighter) curve represents
        the azimuthal angle $\alpha$ range of 
        $0\deg \leq \alpha < 45\deg$ 
        ($45\deg \leq \alpha < 90\deg$), 
        i.e., closer to the galaxy major (minor) axes.
        The vertical 
        thick, dotted line in light red
        shows the 
        median 0.5\rvir\ of the quiescent galaxies 
        in the stack.
        }
    \label{fig:mg2detfrac_qs_1d} 
\end{figure*}

We also emphasize that our result of 
extended cool gas around quiescent galaxies 
(at least for those 
with \mstar\ $\geq 10^{10}$\msununit,  
Figure~\ref{fig:mg2detfrac_qs_4p})
is not surprising according to 
existing observational studies.  
Quasar absorption-line studies have measured
a high incidence rate of metal enriched cool gas
around quiescent galaxies 
\citep{Thom2012,Werk2013,Huang2020}
and massive LRGs
\citep{Gauthier2009,Huang2016,Chen2018,Berg2019,Zahedy2019}, 
and the cool gas mass of LRGs 
is even comparable to that of
$\sim$$L^*$ star-forming galaxies \citep{Zahedy2019}.
\citet{Nelson2020} recently used \texttt{TNG50}
to analyze the cool CGM around 
$z\sim0.5$ massive galaxies 
analogous to the observed LRGs.
They showed that the cool gas mass 
increases with halo mass,
which again implies that LRGs and 
quiescent galaxies do not lack cool gas.
They found that 
the cool gas takes the form of thousands of 
$\sim$kpc-size, thermally underpressurized
clouds dominated by magnetic pressure.
This led to their conclusion that 
magnetic fields possibly influence
the formation and the morphology of 
individual clouds of the cool CGM.

It is also worth noting that while 
our result suggests that the \mgII\ gas around  
quiescent galaxies is potentially
less ``disky'' than that around their
star-forming galaxy counterparts,
it is not true that all of the simulated
quiescent galaxies
lack a rotating \mgII\ gas structure.
A few of them have ``disky'' \mgII\ gas,
even though this is less common 
compared to star-forming galaxies.
While a ``disky'' gas structure 
may naively be unexpected for quiescent galaxies,
especially because 
the majority of quiescent galaxies
are elliptical and lenticular galaxies
\citep{Hubble1936,Bernardi2010},
a recent observational work studied 
local quiescent galaxies and 
found a surprising large reservoir 
of cold, rotating \hI\ gas
similar to that around star-forming galaxies
\citep{Zhang2019}.
These authors suggested that the galaxies 
are quenched not because of the lack of gas in general, 
but because of the 
reduced molecular gas content, 
lower star formation efficiency, 
and/or lower dust content
compared to the star-forming galaxies.


\subsection{Is there a cutoff radius 
for the \mgII\ gas to be observationally detected?}
\label{ssec:selfsim}

For both star-forming and quiescent galaxies,
the \mgII\ detection fraction maps
clearly indicate that the \mgII\ gas 
around higher mass galaxies has 
a larger physical extent (i.e., in pkpc).
Comparing the detection fraction contours
with the dashed circles representing 0.5\rvir\ 
(Figures~\ref{fig:mg2detfrac_sf_16p} and
\ref{fig:mg2detfrac_qs_4p})
still shows that higher mass galaxies 
have a more extended \mgII\ gas distribution 
relative to the halo size,
but the difference is less drastic;
also see the Appendix,
where we show the detection maps with pixels 
scaled with \rvir.
This implies that the \mgII\ gas distribution
around massive galaxies is intrinsically more extended.

Observational work has also demonstrated that
the \mgII\ gas distribution depends
on the mass and size of the host galaxy or halo.
Quasar absorption-line studies often find that
the strength of the \mgII\ absorption system
(measured by the equivalent width (EW))
decreases with increasing 
impact parameter $b$ of the quasar sightline,
but the data points around this relation 
show a large scatter 
\citep{Chen2010,Nielsen2013_ii,Huang2020}.
A mass segragation is also
observed in the EW-$b$ relation;
at large impact parameters,
the \mgII\ systems are detected around galaxies
with higher masses \citep{Churchill2013}.
On the other hand, 
plotting \mgII\ EW vs.~$b/$\rvir\
reduces the scatter compared to that of EW vs.~$b$
and improves the statistical significance 
of the fit
\citep{Churchill2013magiicat,Huang2020},
and plotting against $(b/r_\mathrm{vir})^2$
removes the mass segregation
\citep{Churchill2013,Churchill2013magiicat}.
These results suggest that 
the circumgalactic \mgII\ gas distribution
scales with halo mass and radius,
and Churchill \etal~also find that the
majority of the \mgII\ gas resides within 
$b \lesssim 0.3$\rvir.  
The mass segregation observed matches with 
our result that \mgII\ gas 
is more extended around higher mass galaxies.
However, the \mgII\ gas around 
the \EAGLE\ galaxies
extends beyond 0.3\rvir.
As seen from the \mgII\ detection maps in 
Figures~\ref{fig:mg2detfrac_sf_16p} and
\ref{fig:mg2detfrac_sf_rvir},
although most of the detectable \mgII\
around \logmstarmsun\ $< 10$ galaxies
lies within 0.3\rvir,
for \logmstarmsun\ $\geq 10$ galaxies,
the 50\% detection fraction
extends further than 0.3\rvir.

The extended \mgII\ distribution
around \EAGLE\ galaxies of higher mass
also explains the trend of 
the \mgII\ mis-assignment fraction 
in Section~\ref{sec:mg2misassign}.
At a fixed impact parameter,
the \mgII\ mis-assignment fraction
is higher for a less massive galaxy
compared to that for a more massive galaxy,
implying that the \mgII\ gas selected by  
the \deltavlos\ $=500$\kms\ window
is more likely to lie outside of \rvir\
of the lower mass galaxy.
This can be naturally explained by 
\mgII\ spatial extent scaling with halo size,
because a fixed impact parameter in pkpc
represents a larger fraction relative 
to the halo size for a lower mass galaxy.
In fact, instead of selecting the \mgII\ gas 
within a fixed velocity window 
of \deltavlos\ $=500$\kms, 
we have explored using a window 
that scales with the halo mass and size
and then recalculated the \mgII\ mis-assignment fraction.
We used \deltavlos\ $= 2$\halovc, 
where \halovc\ represents the halo circular speed 
\halovc\ $= \sqrt{G M_\mathrm{vir}/r_\mathrm{vir}}$.\footnote{
    We apply the multiplicative factor of two
    such that \deltavlos\ $\approx 500$\kms\ 
    for a $M_\mathrm{vir} \approx 10^{12}$\msununit\
    galaxy.
    }
Comparing the recalculated 
\mgII\ mis-assignment fractions to those
with \deltavlos\ $=500$\kms\
shows negligible differences
for the galaxies in mass bins of 
\logmstarmsun\ $\geq 10$,
whereas those in 
$9 \leq$ \logmstarmsun\ $< 9.5$
($9.5 \leq$ \logmstarmsun\ $< 10$)
show a percentage decrease 
of $\lesssim35$\% (20\%)
relative to the original results obtained from
the $\pm500$\kms\ window.
This is not surprising,
because 2\halovc\ of a lower mass galaxy
covers a smaller velocity range than the 
fixed 500\kms\ window.
While this seems to suggest that observational studies
should use a scalable window while associating 
the \mgII\ gas with target galaxies, 
neither \rvir\ nor \mvir\ 
are measured directly from observations.


\subsection{Identifying cirumgalactic gas around galaxies:  where does the CGM end?}
\label{ssec:dis_misassign_mg2}

The CGM has been defined as the 
gas roughly within \rvir\ and outside of the 
intersellar medium of galaxies 
\citep[e.g.,][]{Tumlinson2017}.
Many observations of the CGM have been
conducted through quasar absorption-line studies,
but associating the absorption system with a
host galaxy and determining whether the absorbing gas
is circumgalactic is not straightforward.
First, observers do not know
where the absorbing gas lies along the sightline
in 3D-space.
Second,
the \rvir\ (and \mvir) of observed galaxies 
is highly uncertain.
Typically, determining the \rvir\ of an observed galaxy
requires the galaxy stellar mass 
(deduced from galaxy photometry)
and the stellar mass-halo mass relation,
which is model-dependent and
has a large intrinsic scatter
\citep[e.g.,][]{Behroozi2013}.
As a result, observations typically associate 
an absorption system with a host galaxy 
if it is at a small projected separation 
from the quasar sightline 
and has a similar redshift 
as the absorption system.
The latter is typically defined using a fixed  
LOS velocity window 
(e.g., \deltavlos $=500$\kms, 
Section~\ref{sec:mg2misassign}).

Where the CGM ends is a
topic of ongoing discussion
\citep{Shull2014}.
If we assume \rvir\ sets the boundary 
of the CGM, then 
the \mgII\ mis-assignment fraction in
Figure~\ref{fig:mg2_misassignment} shows
how often the \mgII\ gas detected at a 
certain impact parameter comes from 
gas outside of \rvir\
but is selected by the
\deltavlos\ $\leq 500$\kms\ criterion.
This rate of mis-assignment is significant.
For example, for a sightline 
at an impact parameter of 100 kpc,
80\% (6\%) of the times \mgII\ is mis-assigned
for a galaxy with
$9 \leq$ \logmstarmsun\ $ < 9.5$
($10 \leq$ \logmstarmsun\ $< 10.5$).
This raises a warning flag
for observational studies of the CGM,
because observers often select gas 
around galaxies using the \deltavlos\ window.
The mis-assigned \mgII\ gas contaminates
the corotation signal
and leads observers to underestimate
the spatial extent of the corotating gas
(Section~\ref{ssec:obs_corot_rvir}).
This also implies that the mis-assigned gas
affects the study of gas kinematics 
in general,
including the Doppler shift and the velocity spread
measured in quasar sightlines.
The mis-assigned gas may increase
the width of the existing velocity component or
create additional velocity components,
depending
on the velocity difference
of gas inside (if detected) and outside of \rvir\
and the spectral resolution of the absorption spectra.

Ultimately,
the issue of ``mis-assigning'' the host galaxy
of the detected \mgII\ gas, 
or circumgalactic gas in general,
originates from the question of what defines the CGM.
For example, 
whether the CGM should be defined using a spatial 
boundary, e.g., a sphere with radius \rvir, 
or perhaps defined by the gas kinematics, 
e.g., the gas should be bound or 
selected using a velocity window.
How to define the CGM can also depend
on the objective of the study.  
For example,
the \rvir\ boundary is sufficient
for studying the angular momentum of 
the cool CGM and how it grows the disk,
because the disk-CGM interface lies well within \rvir.
But to understand gas recycling 
and the chemical evolution of the CGM,
it is necessary to include gas outside of \rvir,
because this gas will eventually be (re-)accreted
and change the metal content of the CGM.
In any case,
it is important to realize the potential bias
of using any criterion of defining 
the extent of the CGM,
associating absorption systems with 
target galaxies while studying
the circumgalactic gas properties,
and comparing results from observational measurements
with those from cosmological simulations.
Note that zoom-in cosmological simulations 
may not even model a volume sufficiently large 
to cover all the mis-assigned gas that 
falls within the velocity window.

\section{Conclusion}
\label{sec:conclusion}

\mgII\ gas has been widely studied in 
circumgalactic observations  to characterize
the properties of the cool, $\sim10^4$ K CGM.  
In this paper, we used the 
high-resolution \EAGLE\ (25 Mpc)$^3$
cosmological simulation
to analyze the \mgII\ gas around $z\approx0.25$ galaxies.
We focused on the \mgII\ morphological and 
rotation structures and examined how they vary
with galaxy properties.
Because observers often select
the \mgII\ gas around target galaxies
using a LOS velocity cut,
we explored how often  
a LOS velocity window of \deltavlos\ $= 500$\kms\
selects \mgII\ gas outside of \rvir\ 
of the target galaxy.  
We discussed how this mis-assigned \mgII\ gas 
affects circumgalactic \mgII\ gas analyses 
in sightline studies.

We found that the \mgII\ gas around star-forming 
galaxies neither has a spherical distribution
nor resides in a thin disk
but has an axisymmetric structure.
Over half of the galaxies have detectable
rotating \mgII\ gas 
20 pkpc from the midplane.  
The picture of an axisymmetric rotating structure
also explains the azimuthal dependence of the 
corotating \mgII\ gas detection.
The corotating gas is less frequently detected
near the projected galaxy minor axes,
which can be explained by 
winds and accretion from preferred directions.
A similar rotating structure 
is less commonly found 
in our small sample of simulated quiescent galaxies.
This potentially suggests that 
the \mgII\ distribution around quiescent galaxies
is generally less ``disky'' and more isotropic.
Nevertheless, 
for both star-forming and quiescent galaxies,
the \mgII\ gas is more extended  
around galaxies with higher masses,
both in terms of the physical size (in pkpc)
and relative to the halo virial radius.

The picture of an axisymmetric rotating \mgII\ structure
around star-forming galaxies
provides support to the interpretation
of the circumgalactic absorption observed 
in quasar sightlines.
These observations detected \mgII\ gas that corotates
with the galaxy disks,
but reproducing the broad \mgII\ linewidth
required a rotating structures 
of tens of kiloparsec thick instead of a thin disk
\citep{Steidel2002,
Kacprzak2010,Kacprzak2011ApJ,Ho2017,HoMartin2020}.
Our results demonstrate that 
thick \mgII\ rotating structures exists,
which plausibly represents the \mgII\ gas structure
probed by the quasar sightlines that detected 
\mgII\ corotation.
Our description of the axisymmetric rotating \mgII\ gas 
also agrees with
the recent \illustrisTNG\ simulation
result of a circumgalactic angular momentum study,
which suggested a cylindrically symmetric CGM
\citep{DeFelippis2020}.
We also noted that
there exist nearby disks
with \hI\ extra-planar gas rotating 
and extending 20 kpc from the disk midplane.  
A future project can use simulations
to examine whether the
\mgII\ gas resembles a scaled-up \hI\ gas structure
and use multi-component disk models 
to analyze the \mgII\ gas in the same way as 
observational studies analyze the \hI\ gas.

Since circumgalactic absorption studies 
often use a LOS velocity cut 
to select gas associated with galaxies,  
we explored how often adopting a
$\pm 500$\kms\ LOS velocity cut
includes \mgII\ gas physically outside of \rvir,
i.e., mis-assigned \mgII\ gas.  
We characterized the \mgII\ mis-assignment fraction
as a function of impact parameter
around host galaxies of different properties
(Figure~\ref{fig:mg2_misassignment}),
and Table~\ref{tb:fitparam_fmis} provides
the fitted parameters for the analytical function
describing the relation.
This provides an estimate for observers {of
how likely it is that the
\mgII\ gas detected in a sightline
actually comes from 
outside of \rvir\ of the target galaxy 
with known stellar mass.
For example, 
at an impact parameter of 100 pkpc,
the $\pm500$\kms\ velocity cut around galaxies
with stellar masses of 
$10^9$-$10^{9.5}$ \msununit\ 
($10^{10}$-$10^{10.5}$ \msununit)
selects detectable \mgII\ gas beyond \rvir\ 
80\% (6\%) of the time.
It would also be interesting to characterize
this \mgII\ mis-assignment issue using other 
cosmological simulations and compare 
with our results,
so that observers can be better informed regarding
how this issue
affects their circumgalactic measurements.
In particular, 
we demonstrated, according to our simulation,
that not only does the mis-assigned \mgII\ gas
increase the \mgII\ detection fraction 
especially at large impact parameters ($\gtrsim 80$ pkpc),
the mis-assigned \mgII\ gas also reduces
the frequency of detecting corotating \mgII\ gas
at impact parameters $\gtrsim0.25$\rvir.
This will lead observers to deduce 
a smaller extent for the corotating gas structure.
Hence, the issue with the mis-assigned \mgII\ gas
raises potential concerns regarding 
the interpretations of 
the circumgalactic gas measurements.
It is important to realize the potential bias
of using different methods to
identify the circumgalactic gas around galaxies
and comparing results
from observations and cosmological simulations.

\vspace{2em}

We thank the referee for 
comments that improved the manuscript.
We} thank Max Gronke, Kim-Vy Tran, 
and Anshu Gupta for insightful discussions and comments
that improved this work.
We also thank Peter Mitchell, Nastasha Wijers, 
and Sylvia Ploeckinger for the early discussions
on this work,
and we greatly appreciate their technical support
with implementing the ionization tables and 
testing the code of making projection maps.
This work is partially supported by 
the National Science Foundation under AST-1817125.
This work used the DiRAC@Durham facility 
managed by the Institute for Computational Cosmology 
on behalf of the STFC DiRAC HPC Facility
(www.dirac.ac.uk). 
The equipment was funded by BEIS capital funding 
via STFC capital grants ST/K00042X/1, ST/P002293/1,
ST/R002371/1 and ST/S002502/1, Durham University and 
STFC operations grant ST/R000832/1. 
DiRAC is part of the National e-Infrastructure.
%

\bibliography{master_eagle20}

\appendix

\counterwithin{figure}{section}

\section{\mgII\ Detection fraction maps
with pixels scaled with \rvir}

In Section~\ref{ssec:morphology},
we explained that for higher mass galaxies,
not only does the \mgII\ gas have a larger radial extent
in terms of its physical size 
(i.e., in pkpc),
but also relative to the halo size.
Figure~\ref{fig:mg2detfrac_sf_rvir} and 
\ref{fig:mg2detfrac_qs_rvir} show
the \mgII\ detection fraction for star-forming
and quiescent galaxies, respectively,
projected at $i=90$\deg.
Each pixel is scaled by \rvir\
of individual galaxies (Section~\ref{ssec:2dmap}).
The \mgII\ gas around higher mass galaxies
still extends to larger radii relative to \rvir\ 
compared to the lower mass galaxies.  
For example, for the star-forming galaxies,
50\% of the 
$10 \leq$ \logmstarmsun\ $ < 10.5$ galaxies ``detect''
\mgII\ gas out to $\approx 0.35$\rvir,
in contrast to only $\approx 0.25$\rvir\
for the $9 \leq$ \logmstarmsun\ $< 9.5$ galaxies.

\begin{figure}[thb]
    \centering
    \includegraphics[width=1.0\linewidth]{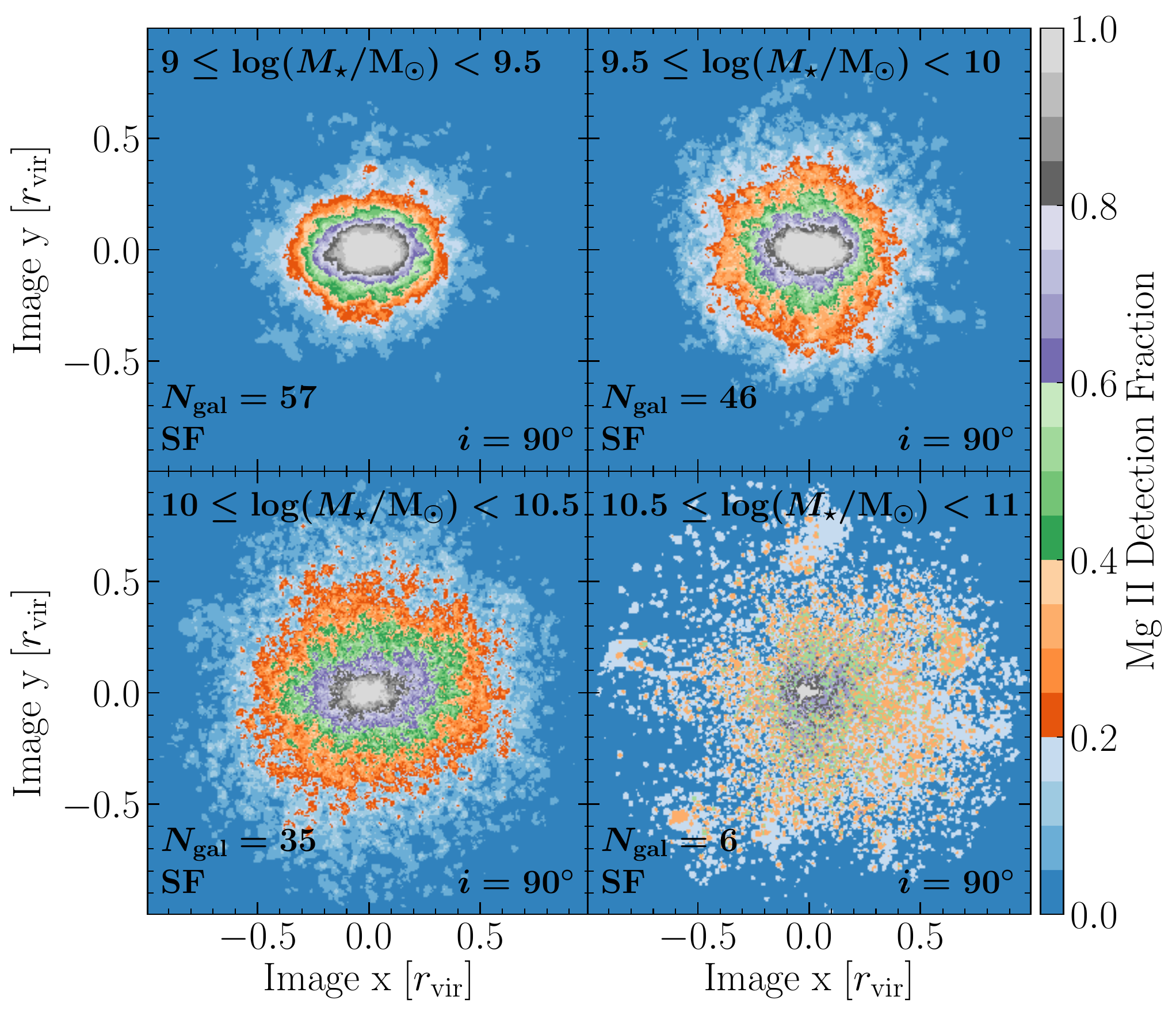}
    \caption{
        \mgII\ detection fraction 
        around star-forming galaxies
        with length parameeters scaled by \rvir.
        The four panels show the stack of 
        star-formaing galaxies
        with different stellar masses 
        (labeled at the top),
        and all galaxies are projected at $i=90$\deg.  
        Compared to lower mass galaxies, 
        the \mgII\ gas around higher mass galaxies
        extends to larger radii relative to \rvir.
        }
    \label{fig:mg2detfrac_sf_rvir} 
\end{figure}

\begin{figure}[thb]
    \centering
    \includegraphics[width=1.0\linewidth]{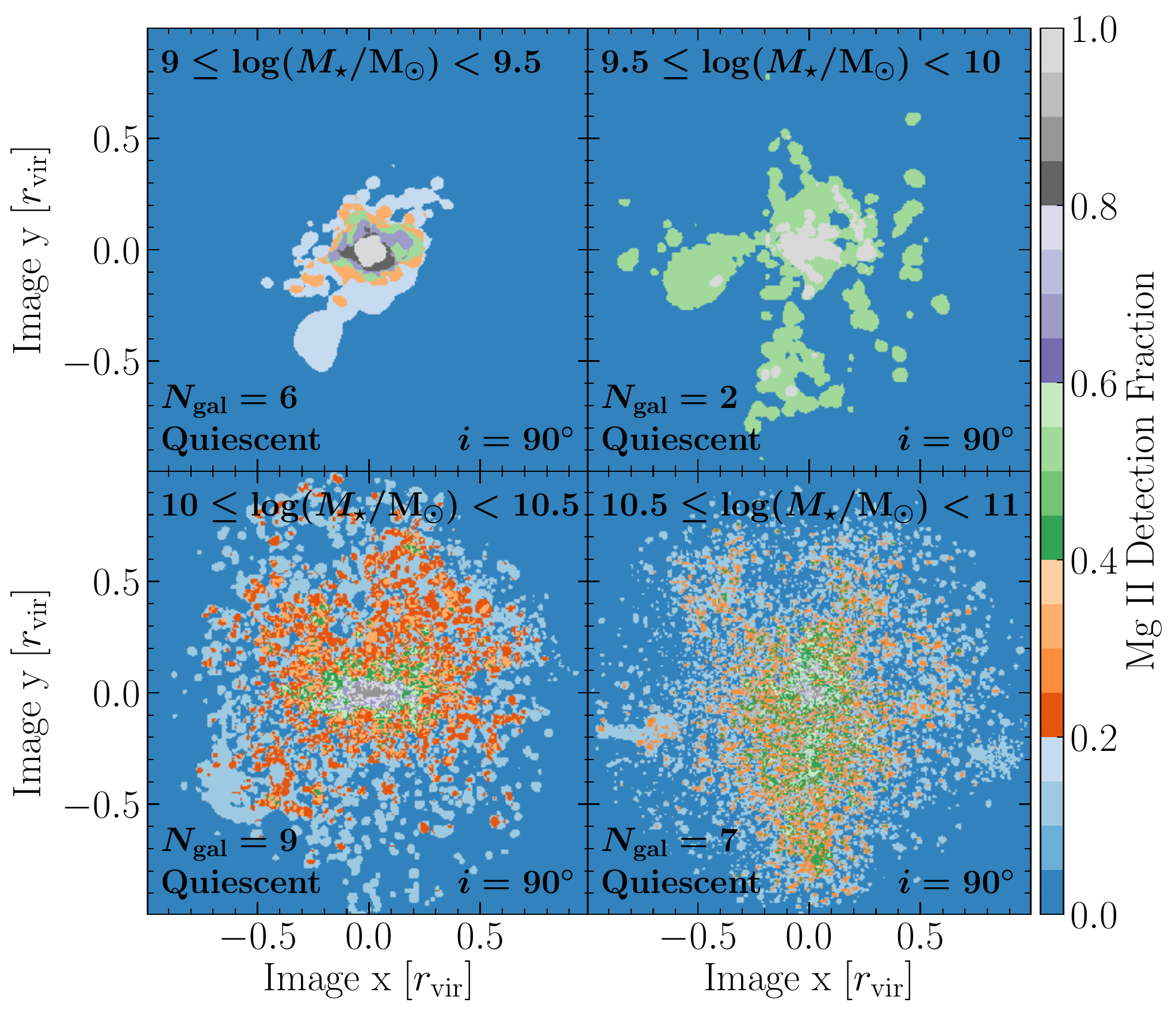}
    \caption{
        \mgII\ detection fraction 
        around quiescent galaxies
        with length parameters scaled by \rvir.
        The four panels have the same arrangement as
        those in Figure~\ref{fig:mg2detfrac_sf_rvir}.
        }
    \label{fig:mg2detfrac_qs_rvir} 
\end{figure}

\section{\mgII\ Detection fraction from
gas selected with \deltavlos\ $\leq 500$\kms}
\label{appsec:mg2detfrac_pm500}

Figure~\ref{fig:detfrac_sf_pm500} shows the 
\mgII\ detection fraction maps
for $i=90$\deg\ star-forming galaxies,
for which the gas is selected to be 
within \deltavlos $= 500$\kms\ of the 
systemic velocities of individual galaxies.  
Comparing these maps with those 
calculated from the \mgII\ gas within \rvir\
(first row of  Figure~\ref{fig:mg2detfrac_sf_16p}), 
the new light blue patches 
near the edge of the maps ($\gtrsim 0.5$\rvir)
indicate an increase in the detection fraction 
from around 0 to $\lesssim 0.1$
if we select the \mgII\ gas 
by \deltavlos\ $\leq\ 500$\kms.

\begin{figure}[thb]
    \centering
    \includegraphics[width=1.0\linewidth]{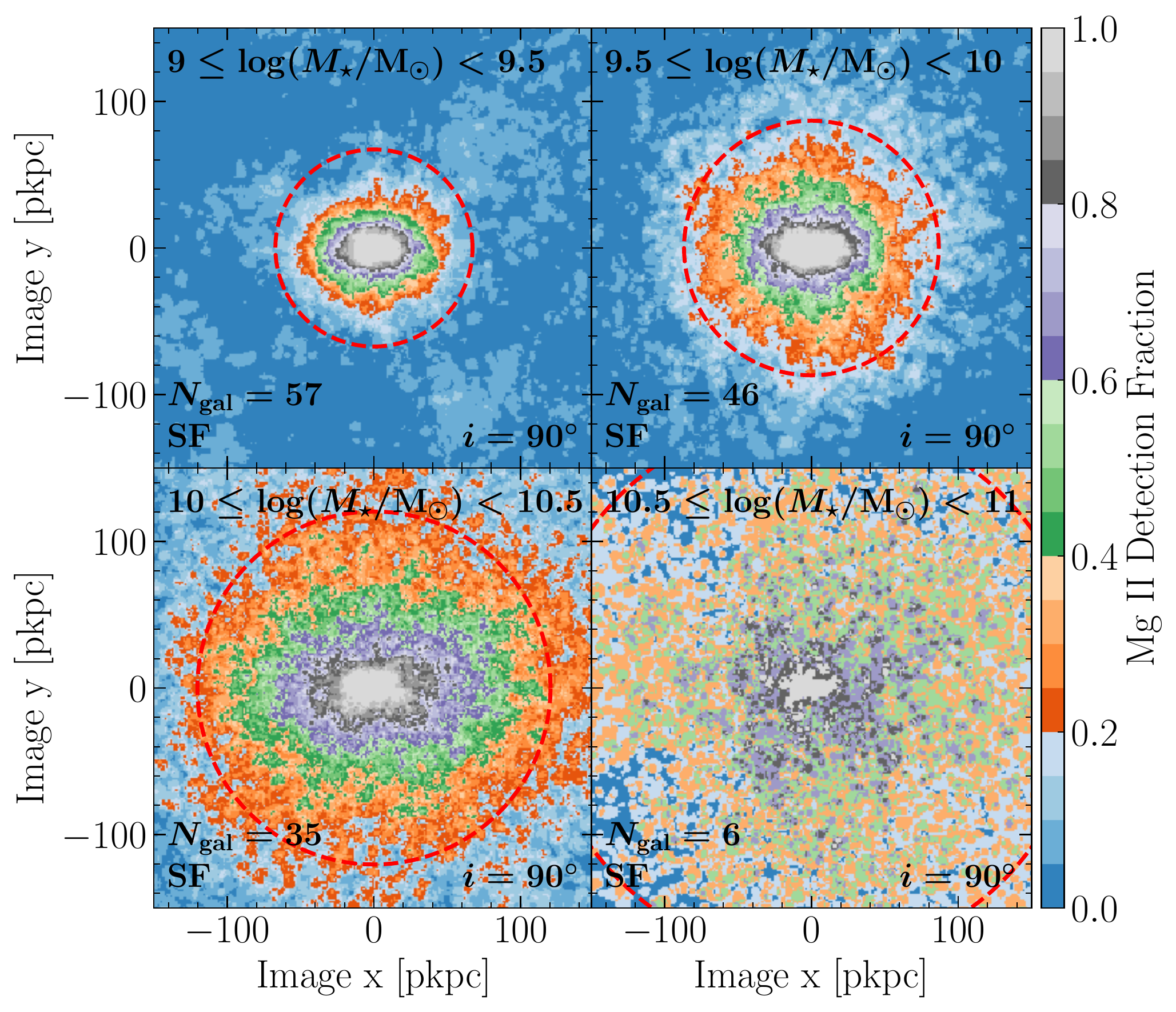}
    \caption{
        Detection fraction of \mgII\ gas 
        within \deltavlos $= 500$\kms\ from 
        the systemic velocities of star-forming galaxies.
        All galaxies are projected at $i=90$\deg.
        This figure is similar to the first row of 
        Figure~\ref{fig:mg2detfrac_sf_16p},
        but instead of selecting the \mgII\ gas 
        enclosed by \rvir,
        here we select the \mgII\ gas using
        the LOS velocity window \deltavlos $= 500$\kms.
        }
    \label{fig:detfrac_sf_pm500} 
\end{figure}

%
%
\end{document}